\pdfoutput=1
\documentclass[fleqn,usenatbib,usedcolumn]{mnras}

\usepackage{graphicx,amssymb,hyperref}
\usepackage{multirow}
\usepackage{color}

\usepackage[fleqn]{amsmath}
\usepackage{abbrev}
\usepackage{siunitx}
\usepackage{comment} 
\usepackage{gensymb}
\usepackage{booktabs,tabularx}
\usepackage{flushend}

\usepackage{nicefrac}
\usepackage{units}
\usepackage{verbatim}

\def\mnras{MNRAS}
\def\apj{ApJ}

\def\simlt{\lower.5ex\hbox{$\; \buildrel < \over \sim \;$}}
\def\simgt{\lower.5ex\hbox{$\; \buildrel > \over \sim \;$}}
\def\etal{{\it et al.}}

\def\D{\mathrm{d}}
\def\rmd{\mathrm{d}}

\def\kpc{\mathrm{\, kpc}}
\def\mpc{\mathrm{\, Mpc}}
\def\msun{\mathrm{\, M_\odot}}
\def\kms{\mathrm{\, km \, s^{-1}}}
\def\cmsg{\, \mathrm{cm^2 \, g^{-1}}}
\def\gyr{ \, \mathrm{Gyr}}

\newcommand{\be}{\begin{equation}}
\newcommand{\ee}{\end{equation}}
\newcommand{\ba}{\begin{eqnarray}}
\newcommand{\ea}{\end{eqnarray}}

\newcommand{\Msun}{\mathrm{M_\odot }}

\DeclareGraphicsExtensions{.pdf,.png}

\title[SIDM in the Bullet Cluster]{What does the Bullet Cluster tell us about self-interacting dark matter?}

\author[A.\ Robertson \etal]{Andrew Robertson\thanks{e-mail: {\tt andrew.robertson@durham.ac.uk}}, 
Richard Massey,
Vincent Eke\\Institute for Computational Cosmology, Durham University, South Road, Durham DH1 3LE, UK\\
}

\begin{document}
\date{ Accepted ---. Received ---; in original form \today.}
\pagerange{\pageref{firstpage}--\pageref{lastpage}} \pubyear{2014}

\maketitle

\label{firstpage}

\begin{abstract}

We perform numerical simulations of the merging galaxy cluster 1E 0657--56 (the Bullet Cluster), including the effects of elastic dark matter scattering. In a similar manner to the stripping of gas by ram pressure, dark matter self-interactions would transfer momentum between the two galaxy cluster dark matter haloes, causing them to lag behind the collisionless galaxies. The absence of an observed separation between the dark matter and stellar components in the Bullet Cluster has been used to place upper limits on the cross-section for dark matter scattering. We emphasise the importance of analysing simulations in an observationally-motivated manner, finding that the way in which the positions of the various components are measured can have a larger impact on derived constraints on dark matter's self-interaction cross-section than reasonable changes to the initial conditions for the merger. In particular, we find that the methods used in previous studies to place some of the tightest constraints on this cross-section do not reflect what is done observationally, and overstate the Bullet Cluster's ability to constrain the particle properties of dark matter. We introduce the first simulations of the Bullet Cluster including both self-interacting dark matter and gas. We find that as the gas is stripped it introduces radially-dependent asymmetries into the stellar and dark matter distributions. As the techniques used to determine the positions of the dark matter and galaxies are sensitive to different radial scales, these asymmetries can lead to erroneously measured offsets between dark matter and galaxies even when they are spatially coincident. 
\end{abstract}


\section{Introduction}

The massive galaxy cluster 1E 0657--56 (the `Bullet Cluster') acts as a dark matter (DM) particle collider, potentially allowing for discrimination between different particle physics models of DM. In particular, limits on the offset between the galaxies and DM associated with the smaller DM halo (the `bullet') as well as limits on the loss of DM mass from the bullet have been used to place constraints on the DM-DM elastic scattering cross-section \citep[][hereafter R08 and K14 respectively]{Randall:2008hs,2014MNRAS.437.2865K}.

Of the myriad of possible DM candidates, the most favoured candidates for the DM particle (e.g. the lightest neutralino in the minimal supersymmetric standard model) typically have only weak non-gravitational interactions. This is not a necessary property of DM, and it was first noted by \citet{Spergel:2000to} that as well as being allowed from a particle physics perspective, DM with a significant cross-section for elastic scattering could have interesting astrophysical consequences. In particular, self-interacting dark matter (SIDM) could alleviate discrepancies between the results of $N$-body simulations of collisionless cold dark matter (CDM) and observations of dwarf galaxies \citep[for a review see][]{Weinberg02022015}. 

While SIDM may not be unique in offering a solution to these `small-scale problems' \citep{2012MNRAS.421.3464P,2012MNRAS.422.1231G,2016MNRAS.457.1931S}, there are numerous DM particle candidates that give rise to scattering between DM particles \citep{1992ApJ...398...43C,2000PhRvD..62d1302B,Kusenko:2001jl,2002PhRvD..66f3002M,2009JCAP...07..004F,2013PhRvL.110k1301T,2014PhRvD..89c5009K,2014PhRvD..89k5017B,2016arXiv160400123W}, so it is an important challenge to try and constrain the cross-section for DM--DM scattering from astrophysical observations, in a bid to constrain the allowed parameter space for DM particle models. If it was found that DM must have a significant self-interaction cross-section this would have a profound effect on particle physics theories of DM, ruling out many of the favoured (and most searched for) DM candidates.

Additional motivation for studying SIDM comes from the detection of separations between the distribution of stars and DM in galaxy clusters \citep{2011MNRAS.415..448W,2014MNRAS.439.2651M,2015MNRAS.449.3393M}. If the offset observed in Abell 3827 \citep{2015MNRAS.449.3393M} is interpreted as resulting from SIDM then it corresponds to an isotropic scattering cross-section of $\sigma / m \sim 1.5 \cmsg$ \citep{2015MNRAS.452L..54K}. While such an offset could potentially arise from an out of equilibrium system, or dynamical effects such as tides or dynamical friction acting differently on the differently distributed stars and DM, offsets of this size appear to be rare in a $\Lambda$CDM Universe \citep{2015MNRAS.453L..58S}.

With this motivation, we revisit (R08, K14) the use of galaxy cluster collisions to constrain the nature of DM. Clusters are useful as their distribution of DM can be probed by both strong and weak gravitational lensing. The relative velocities of DM particles within clusters is also of order $1000 \kms$, two orders of magnitude larger than in dwarf galaxies. Velocity dependent cross-sections can arise naturally in models for SIDM \citep{2009PhRvD..79b3519A,2010PhRvD..81h3522B,2011PhRvL.106q1302L,2012PhRvL.109w1301V,2013PhRvD..87k5007T}, and constraining such models requires a handle on the cross-section at different velocity scales \citep{2016PhRvL.116d1302K}.

The first attempt to use colliding galaxy clusters to constrain the collisional nature of DM \citep{Markevitch:2004dl} found that $\sigma / m < 5 \cmsg$ from limits on the size of any potential offset between the DM and stars in the Bullet Cluster. This constraint, derived from analytical toy models, was improved by R08 who ran $N$-body simulations of Bullet Cluster-like systems with SIDM. Combined with tighter constraints on any DM--galaxy separation \citep{2006ApJ...652..937B}, they found $\sigma / m < 1.25 \cmsg$.

Owing to the high relative velocity of the DM haloes in the Bullet Cluster, DM particles from the bullet that scatter with particles from the main cluster will typically have sufficient energy to escape the potential of the bullet halo, and so the bullet halo would evaporate due to DM self-interactions. The mass to light ratio of the bullet halo is similar to that of the main halo, and if one assumes that this similarity means that less than 23\% of the DM in the inner regions of the bullet halo could have scattered with particles from the main halo then the R08 simulations suggest that $\sigma / m < 0.7 \cmsg$. However, observations of over 200 galaxy clusters \citep{2007A&A...464..451P} have shown that there is significant scatter in the luminosity-mass relation for clusters. Specifically, \citet{2007A&A...464..451P} found that the $r$-band luminosity of clusters was tightly related to the number of galaxies with an $r$-band absolute magnitude of $M_\mathrm{r} \le -20$, but that from the number of galaxies the mass of the cluster could only be predicted with an accuracy of 55\%. This suggests that the significance of the $\sigma / m < 0.7 \cmsg$ result derived in R08 is over-stated, as it assumes little intrinsic scatter in the mass-to-light ratios of clusters.

Since the discovery of the Bullet Cluster, other colliding cluster systems have been found, and used to constrain the cross-section for DM scattering. Similar analysis to that performed on the Bullet Cluster places limits of $\sigma / m < 4 \cmsg$ from MACS J0025.4-1222 \citep{Bradac:2008gw}, $\sigma / m < 3 \cmsg$ from Abell 2744 \citep{2011MNRAS.417..333M}, and $\sigma / m < 7 \cmsg$ from DLSCL J0916.2+2951, the `Musket Ball Cluster' \citep{2012ApJ...747L..42D}.

K14 pointed out that during galaxy cluster collisions, DM particles preferentially collide along the merger axis, and that these systems could be used to determine not just the cross-section for DM scattering, but its angular dependence. In particular, they showed that the resulting distribution of DM is different for the case of short-range, contact interactions (for which the scattering is isotropic), compared to long-range interactions, where there is a preference for low scattering angles, and particles can undergo many small momentum transfer collisions.

Compared with systems undergoing major mergers, clusters undergoing minor mergers with large mass ratios are ubiquitous. \citet{2015Sci...347.1462H} found 30 such clusters, with a total of 72 pieces of substructure. By looking at the position of the DM substructure relative to the position of the corresponding stars and gas, they placed limits of $\sigma / m < 0.47 \cmsg$ for the DM elastic scattering cross-section.

In this work we choose to focus on the Bullet Cluster, as the gas morphology and the lack of line-of-sight velocity difference between galaxies from the two clusters implies that the collision has taken place with little impact parameter and in the plane of the sky \citep{2002A&A...386..816B}. In addition to this simple geometry, X-ray observations of the shock front leading the gaseous bullet allow the relative velocity between the two merging clusters to be estimated \citep{2006ESASP.604..723M}. We limit our study to the case of isotropic and velocity-independent cross-sections, focusing on the importance of the method used to extract position estimates from the simulations.


Our paper is structured as follows. In \S\ref{sect:simulations} we discuss our implementation of DM scattering within an $N$-body code, as well as the initial conditions we use for our simulations. In \S\ref{sect:measuring_positions} we discuss different methods for measuring the positions of different components within a merging galaxy cluster, before applying these different methods to our simulations in \S\ref{sect:results}. Finally, we give our conclusions in \S\ref{sect:conclusions}. We use $\Omega_\mathrm{m} = 0.3$, $\Omega_\Lambda = 0.7$, and $H_0 = 70 \kms \, \mpc^{-1}$. At the redshift of the Bullet Cluster ($z=0.296$) $1 \kpc$ corresponds to $0.23$ arcsec.

\section{Simulations}
\label{sect:simulations}

\subsection{Implementation of DM scattering}

We implemented DM scattering on top of the \textsc{gadget}-3 Tree-PM $N$-body code, which is an updated version of the publicly available \textsc{gadget}-2 code \citep{Springel:2005cz}. The scattering was done stochastically using the same algorithm as in \citet{Rocha:2013bo} that they derive from the Boltzmann equation, although we use a top hat kernel rather than a spline kernel. At each time step, the probability for each pair of nearby particles to scatter is calculated, and a random number is drawn to see which particles do actually scatter. This algorithm is similar to that used in other SIDM simulations \citep{2000ApJ...543..514K,Yoshida:2000gn,2001ApJ...547..574D,2011MNRAS.415.1125K,2012MNRAS.423.3740V,Fry:2015jx}, with these algorithms differing in the number of neighbours (or search volume) used to find potential scattering pairs.

\subsubsection{Assumed DM interaction model}

We assume that the particle interactions are fully described by an azimuthally symmetric differential cross-section, defined in the centre of momentum frame of the two particles, which could have some velocity dependence, $\frac{\D^{2}\sigma}{\D\Omega \, \D v}$. Assuming that the angular and velocity dependences of the cross-section are separable, we can consider only an angularly dependent cross-section. The velocity dependence then enters as a normalisation of the total cross-section between pairs of particles, a function of their relative velocity.

From the differential cross-section, $\frac{\D \sigma}{\D \Omega}$, we can calculate the total cross-section as 
\begin{equation}
\sigma = 2 \pi \int_0^\pi \sin \theta \frac{\D \sigma}{\D \Omega} \, \D \theta.
\end{equation}
We can then define the probability that a scattered particle changes direction by an angle in the range $\left[ \theta, \theta + \D \theta \right]$ as 
\begin{equation}
\label{eq:P(theta)}
P(\theta) \, \D \theta =  \frac{1}{\sigma} 2 \pi \sin \theta \frac{\D \sigma}{\D \Omega} \, \D \theta.
\end{equation}
The code can implement velocity and angular-dependent scattering, but for the rest of this paper we will assume that DM scattering is velocity-independent and isotropic, for which $\frac{\D \sigma}{\D \Omega} = \sigma/4 \pi$. Point-like interactions that lead to isotropic scattering result from scattering mediated by a heavy particle, when the momentum exchange in the scattering is significantly larger than the mass of the mediator particle, $m_\phi$, i.e. when $m_\phi \gg (v/c) m_\mathrm{DM}$.

\subsubsection{Scattering rate}

The scattering rate of an individual DM particle $i$, with velocity $\boldsymbol{v_{i}}$, is
\begin{equation}
\Gamma_{i} = \int f(\boldsymbol{v'}) \, \rho \, \frac{\sigma}{m} \, |\boldsymbol{v_{i}} - \boldsymbol{v'}| \, \rmd^{3} \boldsymbol{v'},
\label{Gamma_i}
\end{equation}
where $f$ is the velocity distribution function,\footnote{Here $f$ is normalised such that $\int f(\boldsymbol{v}) \, \D^{3}\boldsymbol{v} = 1$.} $\rho$ the local density, and $\sigma/m$ the cross-section for DM-DM scattering (which could depend on $|\boldsymbol{v_{i}} - \boldsymbol{v'}|$) divided by the DM particle mass. To calculate scattering probabilities in the simulations, $f$ and $\rho$ are estimated from the volume within a distance $h$ of a particle's position. This leads to the scattering rate 
\begin{equation}
\Gamma_i = \sum_j  \frac{\sigma_\mathrm{p} |\boldsymbol{v_{i}} - \boldsymbol{v_{j}}|}{\frac{4\pi}{3}h^{3}},
\label{eq:gamma_i}
\end{equation}
where the sum is over all simulation particles in the volume defined by $h$, and $\sigma_\mathrm{p} \equiv (\sigma / m) \, m_\mathrm{p}$ with $m_\mathrm{p}$ the mass of the simulation particles. Throughout this work, $\sigma$ and $m$ will be the cross-section and mass of individual DM particles, while $\sigma_\mathrm{p}$ and $m_\mathrm{p}$ will be the cross-section and mass of the simulation particles. Astrophysical observables, such as core sizes, are determined by the fraction of particles that are scattered during a process, and so relate to the scattering rate for individual particles. As the scattering rate depends on the product of the number density of particles and the cross-section, and the number density of particles scales inversely as the particle mass, the cross-section of our simulation particles must scale with the simulation particle mass, such that
\begin{equation}
\label{eq:sig_p}
\sigma_\mathrm{p} =  \left(\frac{\sigma}{m}\right) m_\mathrm{p}.
\end{equation}

From equation~\eqref{eq:gamma_i}, the probability of two particles, $i$ and $j$, separated by a distance less than $h$, scattering within the next time step, $\Delta t$, is given by   
\begin{equation}
\label{eq:P_ij}
P_{ij}= \frac{\sigma_\mathrm{p} |\boldsymbol{v_{i}} - \boldsymbol{v_{j}}| \, \Delta t}{\frac{4\pi}{3}h^{3}}, 
\end{equation}
where for velocity-dependent cross-sections, $\sigma_\mathrm{p}$ would be a function of $|\boldsymbol{v_{i}} - \boldsymbol{v_{j}}|$.

In this scattering procedure, $h$ is a numerical parameter that has to be chosen. In \S\ref{sect:SIDMtests} we investigate the effects of changing $h$, using both a fixed $h$ for all particles, as well as a variable $h$ that adapts to the local density. We also test adding a kernel weighting to the scattering probability in equation~\eqref{eq:P_ij}, such that the probability of nearby pairs of particles (separation much less than $h$) is larger than that for particles further apart. We find that using a fixed value of $h$ equal to the gravitational softening length provides correct results, and that kernel weighting has little effect. This is in contrast to SPH (for which kernel weighting and adaptive smoothing lengths are necessary) as the scattering is inherently stochastic, and so it is not important that the scattering probability varies smoothly.

\subsubsection{Scattering kinematics}

If two particles with identical mass, and velocities $\boldsymbol{v_{i}}$ and $\boldsymbol{v_{j}}$ are to scatter, then we first move into the centre of momentum frame, in which the velocities are $\boldsymbol{v_{i}}'$ and $\boldsymbol{v_{j}}' = -\boldsymbol{v_{i}}'$. We use the direction of $\boldsymbol{v_{i}}'$ to define the $z$-axis, from which $\theta$ is measured. Then we draw a random $\theta$ from $P(\theta)$ to determine the polar angle at which the particles scatter, as well as drawing a random number that we convert into an azimuthal angle. With these two angles, the scattering kinematics in the centre of mass frame is completely determined. Finding the momentum transfer in the centre of mass frame, we can then move back to the simulation frame, and apply these momentum kicks.

For the case of isotopic scattering that we consider in this work, the scattering kinematics is particularly simple. The post-scatter velocities are
\begin{align}
\label{eqn:isotropic_velocities}
\begin{split}
\boldsymbol{u_{i}} = \boldsymbol{V} + w \, \boldsymbol{\hat{e}}  \\
\boldsymbol{u_{j}} = \boldsymbol{V} - w \, \boldsymbol{\hat{e}}
\end{split}
\end{align}
where the $\boldsymbol{u}$ are the post-scatter velocities, $\boldsymbol{V} = (\boldsymbol{v_{i}} + \boldsymbol{v_{j}})/2$, $w=|\boldsymbol{v_{i}}-\boldsymbol{v_{j}}| / 2$, and $\boldsymbol{\hat{e}}$ is a randomly oriented unit vector.

\subsubsection{Multiple scatters within a time step}

As particle scattering is implemented on a particle by particle basis, it is possible for a particle to scatter more than once in a single time step. While the low rate of particle scattering\footnote{For $\sigma / m = 1 \cmsg$ \citet{2015MNRAS.453.2267R} showed that the average number of scattering events per particle is $\mathcal{O}(1)$ by redshift zero, and so the frequency of particles scattering twice within a single simulation time step is very low.} results in these multiple scatters being infrequent, it is important that they are dealt with in an appropriate way. Because the momentum kick from one scattering event alters the velocities of the particles for any future scattering event, we cannot allow a particle to scatter twice in one time step with the same initial velocity. Instead we arbitrarily order all pairs of particles, and update the particle velocities when we decide two particles will scatter. In this way any future scattering events involving these particles will use the updated velocities, essentially time-ordering the scattering events within one simulation time step.

A further complication arises when running simulations on multiple processors. In order for particles that reside on different processors (which have access to different memory) to scatter, the properties of one of the particles must be exported to the processor on which the other particle resides. To increase parallel efficiency, all of these exports take place simultaneously, and then processors determine if any of their imported particles scatter with their own particles. During this step, it would be possible for a particle that is currently exported to scatter off an imported particle on its own processor. As such, the particle could scatter simultaneously on different processors, which would lead to both scattering events taking place with the same initial velocity for the particle in question. Conserving energy and momentum in each of the two scattering events, and then later combining the two momentum kicks, does not in general conserve energy, and this process could lead to the production of additional kinetic energy in the simulation.

To prevent this, we assign a direction between each pair of processors, and only allow particles to be exported in this direction. We use a constant search radius for all particles, so that the search is symmetric, and particles that would have been exported in the other direction will still have a chance to scatter with particles that they would find there, when those particles are imported. Essentially, particles with a separation less than $h$ that reside on different processors meet each other once and scatter with the probability given in equation~\eqref{eq:P_ij}, while particles on the same processor meet twice, but have half this probability of scattering on each occasion. Assigning a directionality to the particle send/receive process does not completely prevent particles scattering simultaneously on different processors, as it is still possible where particles on three or more domains\footnote{A domain is a region of simulation space that is stored on one processor.} are within a distance $h$ of each other. However, the rate of scattering within a time step is low, and the size of the domains compared to the size of $h$ is large, such that these events are highly unlikely.

We keep a log of all scattering events that allows us to detect these problematic encounters. For three particles drawn from a Maxwell-Boltzmann velocity distribution, the mean change in energy when one particle scatters `badly' from the two other particles, is $\left< \Delta E \right> = \frac{1}{2} \left< \mathrm{KE} \right>$, where $\left< \mathrm{KE} \right>$ is the mean kinetic energy of individual particles. This rises to $\left< \Delta E \right> \approx 0.87 \left< \mathrm{KE} \right>$ when we weight the triplets of particles by the probability of them scattering, as particles with higher relative velocities are more likely to scatter (equation~\eqref{eq:P_ij}).

In the simulations used in the rest of this work only one bad scattering event happened. As the expected change in energy due to a bad scattering is of the order of the kinetic energy per particle in the simulation, a bad scattering event changes the total energy by $\sim$ 1 part in $N_\mathrm{DM}$, where $N_\mathrm{DM}$ is the number of DM particles in the simulation. In our simulations this corresponds to 1 part in $10^7$, making it inconsequential compared to the non-conservation of energy from gravitational forces. With variable time steps, manifest energy conservation is lost \citep{Dehnen:2011gu}, and we find that in our simulations the typical level of energy conservation over the course of a simulation is $\sim$ 1 part in $10^4$.

\subsection{Initial conditions}
\label{sect:InitCond}

In order to draw meaningful conclusions on the properties of DM from a comparison of our simulations to observations, it is important that the simulations do a reasonable job of recreating the Bullet Cluster's observed properties. \citet{2014ApJ...787..144L} performed a large suite of magnetohydrodynamic simulations of the Bullet Cluster, hoping to match a wide range of observational data sets. In order to constrain the 34 parameters required to generate their initial conditions, they required over 1000 simulations, which in our case (looking at the effect of changing the DM-DM scattering cross-section) would have to be done for each cross-section that we investigate. This would be an exceptionally computationally-demanding task, and although complicated, the initial conditions generated are still idealised models for the two clusters, ignoring the effects of mass accretion prior to or during the merger, and without substructure that could be important for matching to the lensing data. Instead of attempting the demanding task of finding optimal initial conditions for each cross-section we investigate, we choose to take a simple idealised model for the system, which provides a reasonable match to key data sets. We then investigate how observables (in particular the offset between DM and galaxies) change as the DM cross-section is varied.

\subsubsection{Density profiles}
\label{subsect:DensProfile}

The main constraints on the total density profiles of the two clusters come from lensing observations. As a first model, we take the best-fitting values from fitting two spherically symmetric \citet*[][hereafter NFW]{1997ApJ...490..493N} mass distributions to weak lensing data, as done in \citet[][hereafter SF07]{Springel:2007bg}. With our assumed cosmology, the best fit values are $r_{200} = 2136 \kpc$, $c=1.94$ and $r_{200} = 995 \kpc$, $c=7.12$, for the main cluster and bullet cluster respectively. Given the redshift of the system at  $z = 0.296$, the masses of the two haloes are then $M_{200} \approx \num{1.5e15} \, \Msun$ for the main cluster and $M_{200} \approx \num{1.5e14} \, \Msun$ for the bullet cluster.\footnote{We define $r_{200}$ as the radius at which the mean enclosed DM density is 200 times the critical density, and $M_{200}$ as the mass enclosed within $r_{200}$. The concentration, $c$, is then $r_{200} / r_\mathrm{s}$, where $r_\mathrm{s}$ is the NFW scale radius.}

The concentration of the main halo derived from weak lensing would place this halo well below the concentration-mass relation derived from observations of galaxy clusters \citep{2015ApJ...806....4M} or from numerical \citep{2012MNRAS.423.3018P,2014MNRAS.441.3359D,2015ApJ...799..108D} or analytical \citep{2015MNRAS.452.1217C} work. SF07 found that with $c=2$ the ram pressure on the gas bullet is not sufficient to strip it away from its DM halo. The observed gas bullet trails its DM by $\sim 100 \kpc$, which they could match by increasing the concentration of the main halo to $c=3$. Making the main halo even more concentrated than this resulted in over-predicting the gas-DM separation, and also lead to the morphology of the bow shock differing from what is observed. We therefore choose to use $c=3$ rather than the weak-lensing derived $c=1.94$ for the main halo in our fiducial model for the collision.

We model the total matter distribution of each cluster with a Hernquist profile \citep{Hernquist:1990tq},
\begin{equation}
\rho(r) = \frac{M}{2 \pi} \frac{a}{r} \frac{1}{(r+a)^3}.
\label{eq:Hernquist}
\end{equation}
These are used because unlike NFW profiles, they have a finite mass and so do not need to be truncated. They also have analytical distribution functions, which allow equilibrium initial conditions to be easily generated, and quantities such as the expected scattering rate within a halo to be calculated analytically.

In order to match an NFW profile to a Hernquist profile we need to define two matching criteria to fix the Hernquist profile's two free parameters. The first of these we take to be matching the normalisation of the density in the central regions, for which $\rho \propto r^{-1}$ for both NFW and Hernquist profiles. We also then match the mass within a radius of $r_{200}$ for the Hernquist profile to that of the NFW profile, making use of the mass within a radius $r$ for a Hernquist profile,
\begin{equation}
M(<r) = M \frac{r^2}{(r+a)^2}.
\label{Mr_Hernquist}
\end{equation}

Enforcing these matching criteria we can find the relationship between the Hernquist parameters, $a$ and $M$, and the NFW parameters, $M_{200}$, $r_{200}$, and $c$:
\begin{equation}
M = M_{200} \frac{(r_{200}+a)^2}{r_{200}^2}
\end{equation}
\begin{equation}
\label{M200-match_a}
a = \frac{r_{200}}{ \sqrt{\frac{c^2}{2 \left[ \ln (1+c) - \frac{c}{1+c}\right]}} - 1}.
\end{equation}
We note that a similar matching procedure is described in the text of \citet{2005MNRAS.361..776S}, but that they match $M_{200}$ of the NFW profile to the total mass, $M$, of the Hernquist profile, resulting in a slightly different formula for $a$.

\subsubsection{Relative velocity of the DM haloes}
\label{sect:relative_vel}

The relative velocity between the two DM haloes in the Bullet Cluster was originally estimated to be $4700 \kms$, as this corresponded to the `shock velocity', the velocity of the shock front relative to the pre-shocked gas \citep{2006ESASP.604..723M}. This large relative velocity would be rare within the context of $\Lambda$CDM \citep{2006MNRAS.370L..38H}, leading to the suggestion of a long-range fifth-force that would result in additional acceleration \citep{2007PhRvL..98q1302F}.

Simulations including gas have since shown that the shock velocity can be considerably larger than the relative velocity between the DM haloes. The pre-shocked gas, which belongs to the main halo, is not at rest with respect to its halo, but is instead moving towards the bullet halo. Additionally, the shock front is not at rest with respect to the bullet halo DM, but moves ahead of it. A discussion of the mechanisms responsible for these effects is available in SF07.

SF07 find that the observed shock velocity can be matched by haloes that collide with a velocity corresponding to infall from infinity. We therefore start our simulations with the cluster centres separated by $4 \mpc$ and with a relative velocity that corresponds to the velocity they would obtain if falling from rest at infinite separation, assuming each halo acts like a point mass.


\subsubsection{Summary of initial conditions}
\label{sect:IC_summary}

Our fiducial model for Bullet Cluster-like initial conditions is two Hernquist profiles, separated by $4\mpc$, and with a relative velocity of $2970 \kms$ along the line joining the two cluster centres. The main halo corresponds to an NFW profile with $M_{200} = \num{1.5e15} \msun$ and $c=3$, while the bullet halo has $M_{200} = \num{1.5e14} \msun$ and $c=7.12$. When converted into matched Hernquist profiles (following the method in \S\ref{subsect:DensProfile}), the masses and scale radii are $M = \num{3.85e15} \msun$, $a=1290 \kpc$, and $M = \num{2.46e14} \msun$, $a=279 \kpc$ for the main and bullet halo respectively.

The mass within each halo is 99\% DM, and 1\% stars, though we use an equal number of DM and star particles ($10^7$ of each). The star particles are distributed as a smooth halo following the DM density. While this is not the case in real galaxy clusters, where stars reside within galaxies, we do this to allow us to more easily identify the location of the stellar component. We also run some simulations including non-radiative gas, which are discussed in \S\ref{sect:including_gas}. The gas initially follows the same density profile as the DM and stars, with the halo mass being 83\% DM, 16\% gas, and 1\% stars. The gas temperature was set so that the gas was in hydrostatic equilibrium, which for the main halo in our fiducial mass model gave a maximum gas temperature of $8.4 \, \mathrm{keV}$, in agreement with the temperature of the pre-shocked gas in the Bullet Cluster \citep{2006ESASP.604..723M}.

\subsubsection{Comparison to other SIDM studies}

\begin{figure*}
        \centering
        \includegraphics[width=\textwidth]{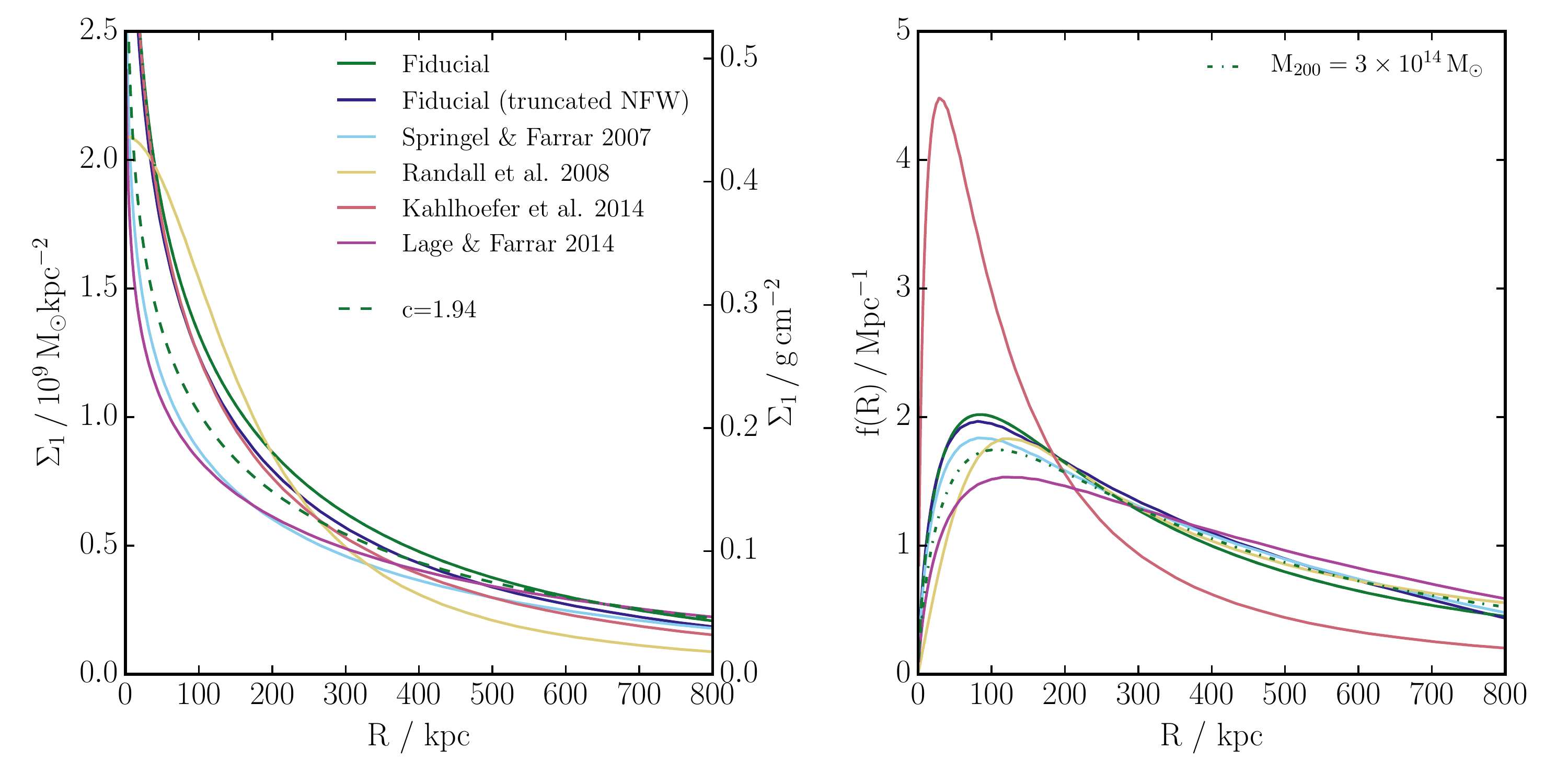}
	\caption{Left panel: the projected mass density through the main halo of the Bullet Cluster, as a function of projected radius. Right panel: the distribution of DM mass at different projected radii in the subcluster of the Bullet Cluster, normalised so that $\int_0^{1 \mpc} f(R) \D R = 1$. Different line styles and colours correspond to different choices for the density profiles. Our fiducial model is described in \S\ref{sect:IC_summary} while two variations to our fiducial model ($c=1.94$ and $M_{200} = \num{3e14} \msun$) are described in \S\ref{sect:varICs}. The Fiducial (truncated NFW) lines are for the underlying NFW profiles that our fiducial model (which uses Hernquist profiles) are matched to, truncated so there is no mass outside of $r_{200}$. Springel \& Farrar 2007 was the fiducial model used in that paper, while Randall et al. 2008 is for the density profiles used in their simulation with $\sigma / m = 1.25 \cmsg$ (their initial conditions were changed slightly for different cross-sections). Kahlhoefer et al. 2014 only simulated one model for the Bullet Cluster, which had a particularly concentrated bullet halo as evident in the right panel. Lage \& Farrar 2014 was the best-fit model found from running over a thousand simulations with different initial conditions and comparing the results to several observational datasets.}
	\label{fig:projected_density_comparison}
\end{figure*}

In Fig.~\ref{fig:projected_density_comparison} we show the density distribution of the main halo and bullet halo from different simulations of the Bullet Cluster. As we are interested in the offset between stars and DM within the bullet halo, the fraction of DM particles from the bullet halo that scatter from a particle in the main halo is an important quantity. We therefore plot the density distributions of the two haloes in a manner that allows us to estimate this fraction. For the main halo, we plot the projected density of DM at different radii, which can be multiplied by the cross-section to get an optical depth for DM scattering. For example, an SIDM particle with $\sigma / m = 1 \cmsg$ passing through the main halo of our fiducial model at a projected radius of $200 \kpc$, where the projected surface density is $\sim 0.15 \, \mathrm{g \, cm^{-2}}$, would have a $\sim 15\%$ chance of scattering off a particle in the main halo.

In the right-hand panel of Fig.~\ref{fig:projected_density_comparison} we plot the fraction of particles at different projected radii within the bullet halo. As the two haloes collide head-on, this is the distribution of projected-radii of the main halo through which bullet halo particles will pass (if we ignore the motion of DM particles within their own halo). We can then use the two panels of Fig.~\ref{fig:projected_density_comparison} to calculate the fraction of particles in the bullet halo that scatter with a particle from the main halo. For our fiducial model with $\sigma / m = 1 \cmsg$, we expect $\sim$ 23 (33)\% of particles from the inner 400 (150)$\kpc$ of the bullet halo to scatter with a particle from the main halo, while for R08 and K14 the numbers are 21 (33)\% and 28 (36)\% respectively. Considering all particles in the bullet halo, the number goes down to 12\% for our fiducial model, in good agreement with the value of 13\% that we get in our simulations (see \S\ref{sect:offsets_methods}).

\subsubsection{Stability of an isolated halo}

\begin{figure}
        \centering
        \includegraphics[width=\columnwidth]{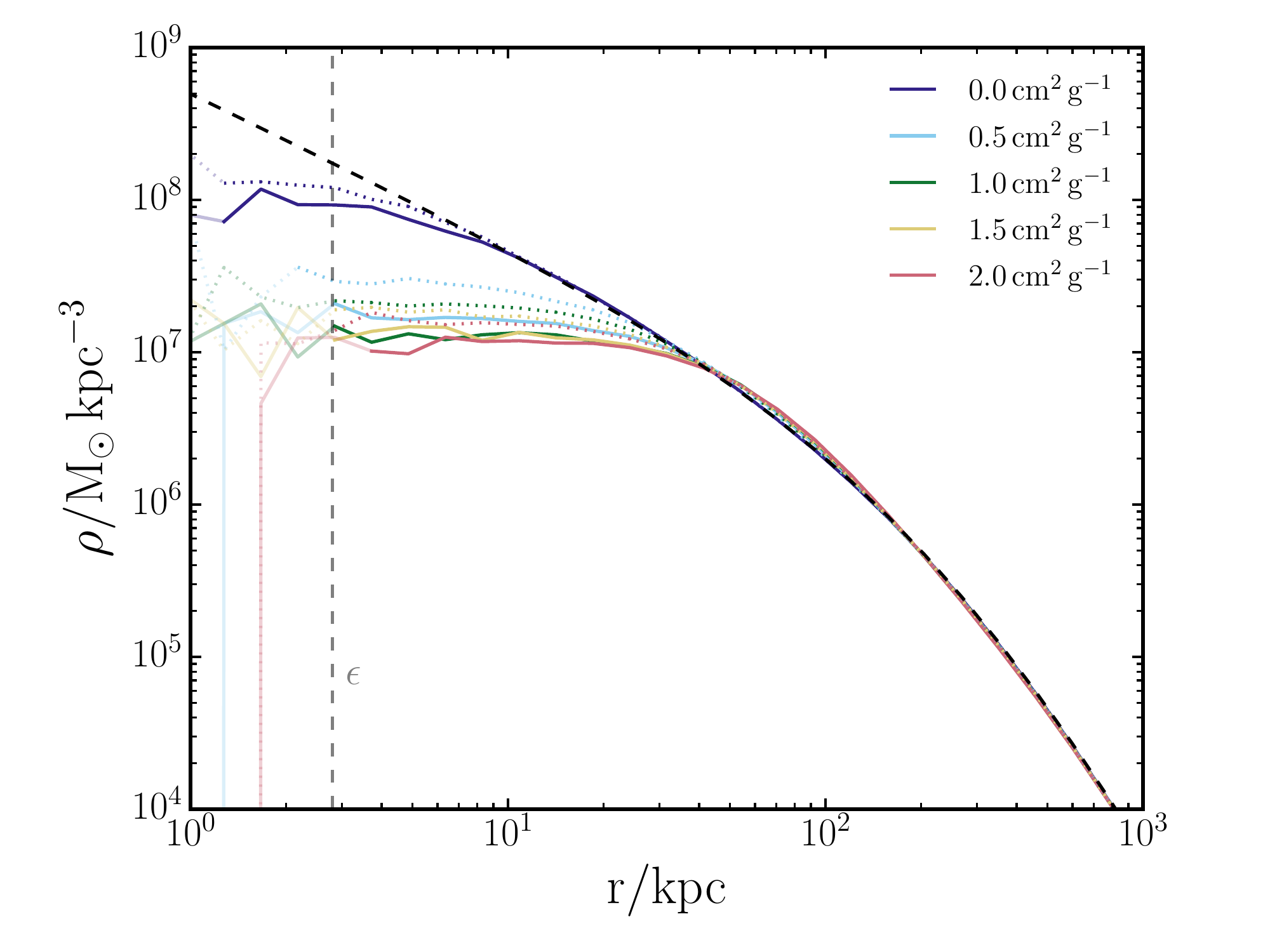}
	\caption{The radial density profile of an isolated halo with collisionless DM as well as SIDM with isotropic cross-sections ranging from $0.5$ to $2 \cmsg$. The dotted lines show the average profile between 1 and 2 Gyr after the start of the simulation, while the solid lines show the period 5--6 Gyr after the start of the simulation. Lines are semi-transparent when the density corresponds to fewer than five particles in a radial bin. The vertical line corresponds to the Plummer-equivalent gravitational softening length, $\epsilon$. For the collisionless DM the initial Hernquist profile (dashed line) is stable, except for the formation of a numerical core with size $\sim 2 \epsilon$ due to gravitational softening.}
	\label{fig:stability_isolated_halo}
\end{figure}

In Fig.~\ref{fig:stability_isolated_halo} we show the density of an isolated Hernquist profile, evolved both with and without DM scattering. The halo shown is the same as the smaller halo in our fiducial model for the Bullet Cluster. With collisionless DM the halo forms a small core with a size $\sim 2 \epsilon$, where $\epsilon$ is the Plummer-equivalent gravitational softening length. The gravitational force between pairs of particles is Newtonian when they are separated by more than $2.8 \epsilon$, but is reduced below this when they are closer, resulting in the formation of small numerical cores in otherwise cuspy haloes.

With SIDM the haloes form much larger cores, due to particles being preferentially scattered out of high density regions. These cores form quickly, and settle to a size that is independent of the DM cross-section, in agreement with \citet{2000ApJ...543..514K}.

Starting our simulations with the cluster centres separated by $4 \mpc$ results in core passage taking place $\sim 1.1 \gyr$ after the simulations begin. During this time the density profiles of the SIDM haloes evolve due to DM scattering, beginning to form constant density cores at their centres. To check that the extent of core formation does not have a large impact on our results, we experimented with a different initial separations between the two haloes. Starting the haloes with a separation of $9 \mpc$, the haloes have evolved for $3.4 \gyr$ before they collide. We found this only had a small impact on our results, changing the best-fit separation between stars and DM (with a scattering cross-section of $1 \cmsg$) from $9.2 \kpc$ to $8.4 \kpc$ at the time of the observed Bullet Cluster in our fiducial model. This change is small compared to the effects discussed in \S\ref{sect:results}.

\subsection{Testing the SIDM implementation}
\label{sect:SIDMtests}

\begin{figure}
        \centering
        \includegraphics[width=\columnwidth]{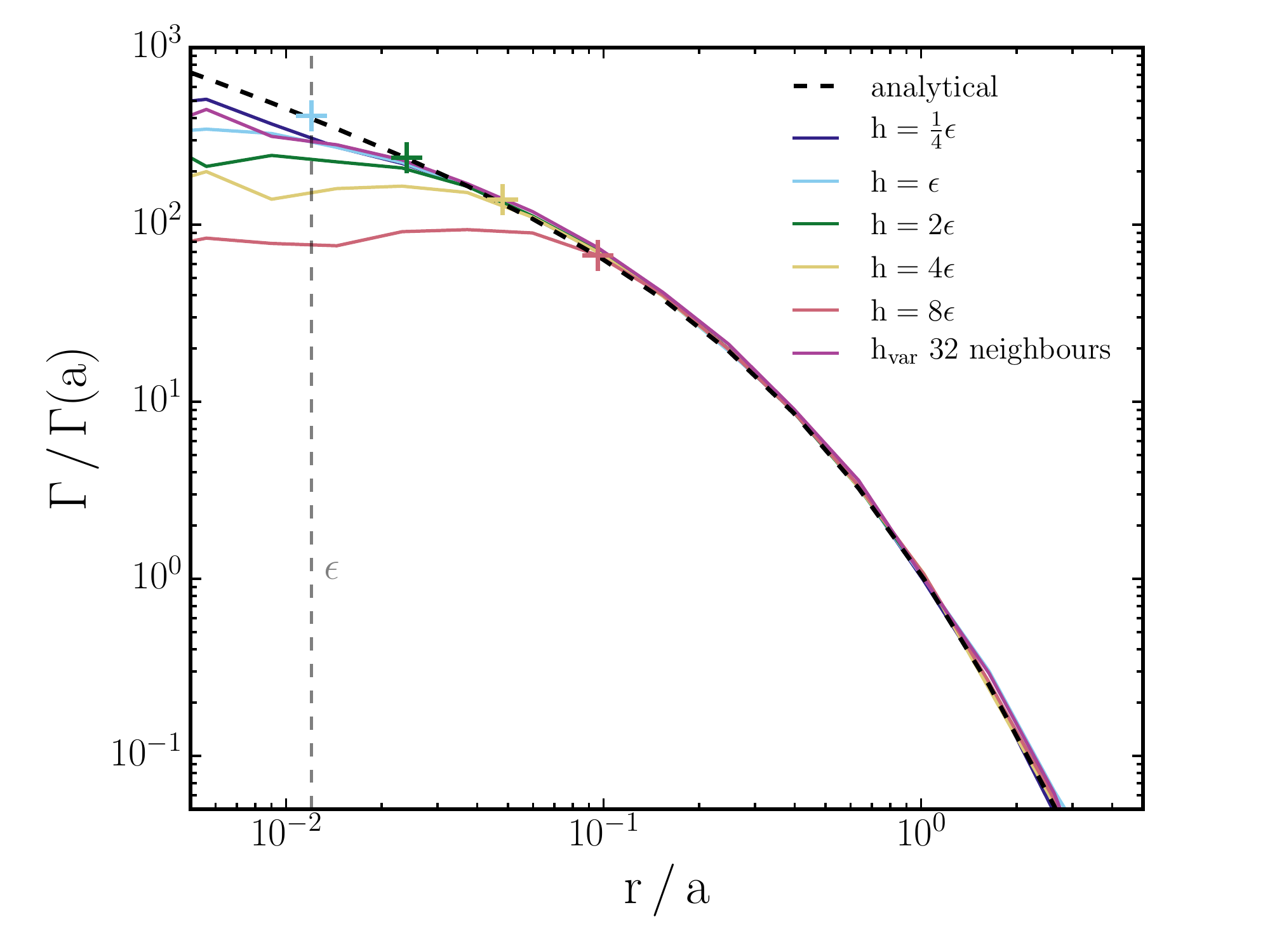}
	\caption{The scattering rate per particle in a Hernquist profile DM halo, plotted as a function of radius. Increasing the size of the search radius used for DM scattering leads to a decrease of the scattering rate in the inner regions of the halo. The results converge for $h < \epsilon$ as the density profile in the simulation forms a numerical core with radius $\sim \epsilon$ due to gravitational softening. All lines used a fixed $h$ except for the $h_\mathrm{var}$ line for which $h$ is varied for each particle to keep 32 neighbours within the search region. For the fixed $h$ lines there are corresponding crosses plotted along the analytical curve at the radius equal to $h$, showing that for $r \lesssim h$ the scattering rate falls below the analytic result.}
	\label{fig:radial_scattering_profile}
\end{figure}

In Fig.~\ref{fig:radial_scattering_profile} we plot the scattering rate per particle in an isolated DM halo with a Hernquist density profile. The halo has a total mass of $M=10^{15} \msun$ and a scale radius $a=1000 \kpc$. The simulation was run for 2.5 Gyr with $10^6$ particles each with a mass $m_\mathrm{p} = 10^9 \msun$, and a Plummer-equivalent gravitational softening length $\epsilon = 12 \kpc$. 

The scattering rate per particle as a function of radius was extracted from the simulations by taking the location of all scatters during the simulation and binning them in logarithmically-spaced radial bins. This was then divided by the time averaged number of particles within the same radial bins to get the scattering rate per particle. 

\subsubsection{Analytical expectation for scattering rates in haloes}

For the analytical calculation of the expected scattering rate per particle, the density and the mean pairwise velocity need to be known. The density distribution is given in equation~\eqref{eq:Hernquist}, while the pairwise velocities can be calculated from the velocity dispersion. Given isotropic velocities following a Maxwell--Boltzmann distribution, the mean pairwise velocity is given by $\left< v_\mathrm{pair} \right> = (4 / \sqrt{\pi}) \sigma_\mathrm{1D}$, where $\sigma_\mathrm{1D}$ is the one-dimensional velocity dispersion. This can be calculated from the density profile and the Jeans equation, which (again assuming an isotropic velocity distribution) gives
\begin{equation}
\begin{split}
\sigma_\mathrm{1D}^2 = &\frac{GM}{12a} \left\{ \frac{12r(r+a)^3}{a^4} \ln \left( \frac{r+a}{r} \right) \right. \\
&- \left. \frac{r}{r+a} \left[ 25 + 52\frac{r}{a} + 42 \left( \frac{r}{a} \right)^2 + 12 \left( \frac{r}{a} \right)^3 \right] \right\},
\end{split}
\label{eq:Hernquist_sig1D}
\end{equation}
for a Hernquist profile.

Integrating over the velocity distribution function in equation~\eqref{Gamma_i} gives the average scattering rate for particles at position $\boldsymbol{r}$,
\begin{equation}
\Gamma(\boldsymbol{r}) = \frac{\langle \sigma \, v_\mathrm{pair}\rangle(\boldsymbol{r})  \rho(\boldsymbol{r}) }{m}.
\label{Gamma_i_r}
\end{equation}
If the DM cross-section is velocity independent then $\langle \sigma \, v_\mathrm{pair}\rangle = \sigma \langle v_\mathrm{pair}\rangle = \sigma (4 / \sqrt{\pi}) \sigma_\mathrm{1D}$, and we can calculate the expected scattering rate per particle at different radii in the halo from equations \eqref{eq:Hernquist} and \eqref{eq:Hernquist_sig1D}. This is shown as the dashed line in Fig.~\ref{fig:radial_scattering_profile}.

\subsubsection{Scattering rates in simulated haloes}

As DM scattering leads to the formation of a cored density profile and also changes the velocity distribution, the scattering rate as a function of radius would not follow the analytical relation once the system has evolved due to self-interactions. To allow for a direct comparison to the analytical result we turn-off the momentum kicks from scattering, such that the scattering algorithm is used to find particles that scatter, but does not actually change the particles' momenta as a result of scattering.

Fig.~\ref{fig:radial_scattering_profile} demonstrates that our code reproduces the correct scattering rate within the halo at all but the smallest radii -- where the scattering rate falls below the analytic prediction. This behaviour is easily understood by noting that the search radius for finding neighbours from which to scatter, $h$, acts as a scale on which the density and velocity distribution are smoothed in the calculation of scattering probabilities. The search radius therefore smooths away the density cusp in the scattering rate calculation leading to decreased scattering rates compared to the true unsmoothed rate. The scattering rate in the simulations drops significantly below the analytic rate only for radii less than $h$, so using a small $h$ is preferred to capture the scattering rate in small high-density regions.

For $h$ smaller than the gravitational softening length, $\epsilon$, the radius within which $\Gamma$ falls below the analytical result ceases to change. This is because there is a core formed in the particle distribution due to gravitational softening, with the core size of the order of $\epsilon$. Pushing $h$ to smaller values than $\epsilon$ therefore cannot recover the analytical result, because the particle distribution is already smoothed on the scale of the gravitational softening.

As using a small $h$ leads to larger probabilities for pairs of particles to scatter (equation~\eqref{eq:P_ij}), smaller time steps must be used to keep these probabilities below 1. We find that setting $h=\epsilon$ allows the usual dynamical time steps to be used, while not excessively smoothing the density in the calculation of scattering rates.

\citet{Rocha:2013bo} found that their scattering algorithm under-predicted scattering rates for small values of $h$ in low-density regions. Specifically, they found the scattering rate dropped below the correct rate when $h \, (\rho / m_\mathrm{p})^{1/3} \lesssim 0.2$, i.e. when $h$ is less than 20\% of the mean inter-particle separation. For $h=0.1 \kpc$ in Fig.~\ref{fig:radial_scattering_profile}, $h$ is $\sim 4\%$ of the mean inter-particle separation at $r=2a$, but the scattering rate still matches the analytical prediction. The result found by \citet{Rocha:2013bo} may be a result of allowing scattering probabilities within a time step to go above 1. This is discussed further, along with some more tests of the SIDM implementation, in Appendix~\ref{App:scattering_tests}.

\section{Measuring Positions}
\label{sect:measuring_positions}

In order to measure the offsets between different components, we first need a definition of position for each of the components. Observationally, the methods used to find the positions of the gas, galaxies and DM are typically all different, and may also be different from the methods used to find the positions in associated simulations. It is therefore important that we understand the effects of changing the method used to find the positions of the various components, in a bid to understand how to best analyse the simulations in order to compare the results with observations.

\subsection{Shrinking Circles}
\label{sect:shrinking_circles}

The \emph{shrinking circles} approach to finding the position associated with a set of discrete points (the simulation particles) is the 2D analogue of the \emph{Shrinking Spheres} approach often used to find density peaks in $N$-body simulations \citep[see e.g.][]{2003MNRAS.338...14P}. All of the particles under consideration are first projected along one axis. Then a circle is drawn. centred on the mean position of all particles, with radius chosen to be the distance between this centre and the most distant particle. The radius is then shrunk by a factor $f$ and a new centre is calculated from the mean position of all particles within the current circle. The radius is shrunk again, and the process continues until the radius of the circle is $R_\mathrm{min}$. The mean position of all particles within this final circle gives the position of this set of particles.

This was the method employed by R08 who used $R_\mathrm{min} = 200 \kpc$. This method clearly only gives one position for a distribution of particles, and so to get the position of both DM haloes from a simulation of the Bullet Cluster the method needs to be run separately on particles belonging to the different haloes, or be started with the circles already shrunk to a size where they only contain one DM peak. 

\subsection{Parametric fits to 2D density maps}
\label{sect:fit_to_kappa}

As an alternative to using shrinking circles to find the positions of the two haloes, we simultaneously fit the projected density map with two profiles that have analytical projected densities. We use 2D projections of Pseudo Isothermal Elliptical Mass Distributions (PIEMDs), which have a 3D density profile
\begin{equation}
\rho(r) = \frac{\rho_0}{(1+r^2 / r_\mathrm{core}^2)(1+r^2 / r_\mathrm{cut}^2)}; \quad r_\mathrm{cut} > r_\mathrm{core}.
\label{PIEMD_density}
\end{equation}
This profile has a core with central density $\rho_0$ and size $r_\mathrm{core}$, outside of which $\rho \propto r^{-2}$ as for an isothermal sphere, until $r \gtrsim r_\mathrm{cut}$ for which the density falls off as $r^{-4}$. This density profile is useful in these SIDM simulations, where the additional free parameter over an NFW or Hernquist profile, allows the cores produced by DM scattering to be well-fitted. The 3D potential and projected-potential are also analytical for this model, making it popular in gravitational-lensing analyses where deflection angles, shears and convergence depend on gradients of the projected potential.

The projected density for a PIEMD is
\begin{equation}
\begin{split}
\Sigma(R) &= 2  \int_R^\infty \frac{\rho(r) r}{\sqrt{r^2 - R^2}} \, \D r \\
&= \Sigma_0 \frac{r_\mathrm{core} \, r_\mathrm{cut}}{r_\mathrm{cut} - r_\mathrm{core}} \left( \frac{1}{\sqrt{r_\mathrm{core}^2+R^2}} + \frac{1}{\sqrt{r_\mathrm{cut}^2+R^2}}  \right),
\end{split}
\label{PIEMD_R}
\end{equation}
where $R$ is the projected radius from the centre of the halo, and 
\begin{equation}
\Sigma_0 = \pi \rho_0  \frac{r_\mathrm{core} \, r_\mathrm{cut}}{r_\mathrm{cut} + r_\mathrm{core}}.
\label{Sigma_0}
\end{equation}

As described in \citet{1993ApJ...417..450K}
, the axially symmetric projected density profile in equation~\eqref{PIEMD_R} can be made elliptical by substituting $R \to \tilde{R}$, where
\begin{equation}
\tilde{R}^2 = \frac{\tilde{x}^2}{(1+\epsilon)^2} + \frac{\tilde{y}^2}{(1-\epsilon)^2},
\label{R_prime}
\end{equation}
and $\tilde{x}$ and $\tilde{y}$ are the spatial coordinates from the centre of the halo, along the major and minor projected axes of the halo respectively. The ellipticity of the halo is defined as $\epsilon = (a-b)/(a+b)$ where $a$ and $b$ are the semi-major and semi-minor axes. Along with $\epsilon$ there is an additional parameter $\phi$ that describes the angle between the $\tilde{x}$-axis and the $x$-axis, i.e. the position angle of the major axis of the halo relative to our coordinate system $(x,y)$.   

We find the 2D positions of the DM haloes by simultaneously fitting two PIEMDs to the total projected DM density. We first discuss the case of fitting the distribution to a single isolated halo, the progression to two haloes then being relatively straightforward.

Each halo is described by seven parameters: the coordinates of the centre $(X,Y)$, the central density $\rho_0$, the core radius $r_\mathrm{core}$, the outer radius $r_\mathrm{cut}$, the ellipticity $\epsilon$ and the position angle $\phi$. The distribution of simulation particles is split into evenly sized bins, generating the data map, $d_{ij}$, to which we find the best-fitting parametric model. Given values for the seven parameters that describe a PIEMD, the surface density can be calculated at each bin using $\Sigma(R)$ from equation~\eqref{PIEMD_R} and using $R \to \tilde{R}$ calculated as the distance between the centre of each bin and the halo centre $(X,Y)$ transformed according to equation~\eqref{R_prime}. This would more accurately be done by integrating $\Sigma(x,y)$ over the area of the bin. As the density is roughly constant for $\tilde{R} < r_\mathrm{core}$, and our bin size used is smaller than the core radii found, the variation of $\Sigma$ across any individual bin is small, and the mean surface density within a bin is well approximated by the surface density at the bin centre.

The model map, $m_{i}$, is the expected number of particles in each bin given the current parameter values, $\theta$. This is simply the surface density at the bin position multiplied by the bin area, and divided by the mass of the simulation particles.

Once we have a data map and a model map, we can calculate the probability of getting our data map given the model map (i.e. the likelihood). For a bin with a given model value, we expect the data value to be Poisson distributed with the expectation value equal to the model value. The likelihood is the product over all map bins of the probabilities of obtaining each data value given the model value: 
\begin{equation}
\mathcal{L}(\theta=\{ X, Y, \rho_0, r_\mathrm{core}, r_\mathrm{cut}, \phi, \epsilon \} ) = \prod_{i} \frac{m_{i}^{d_{i}} \operatorname{e}^{-m_{i}}}{d_{i}!}.
\label{L_kappa}
\end{equation}

We can combine this likelihood function with a set of priors to calculate posterior probabilities for the parameters. This is done using \textsc{emcee} \citep{ForemanMackey:2013io}, a \textsc{python} implementation of the affine-invariant ensemble sampler for Markov chain Monte Carlo (MCMC) proposed by \citet{Goodman:2010et}. We choose flat priors for $X$, $Y$, $r_\mathrm{core}$ and $r_\mathrm{cut}$, with a prior on $\rho_0$ that is flat in log-space.

\begin{figure*}
        \centering
        \includegraphics[width=\textwidth]{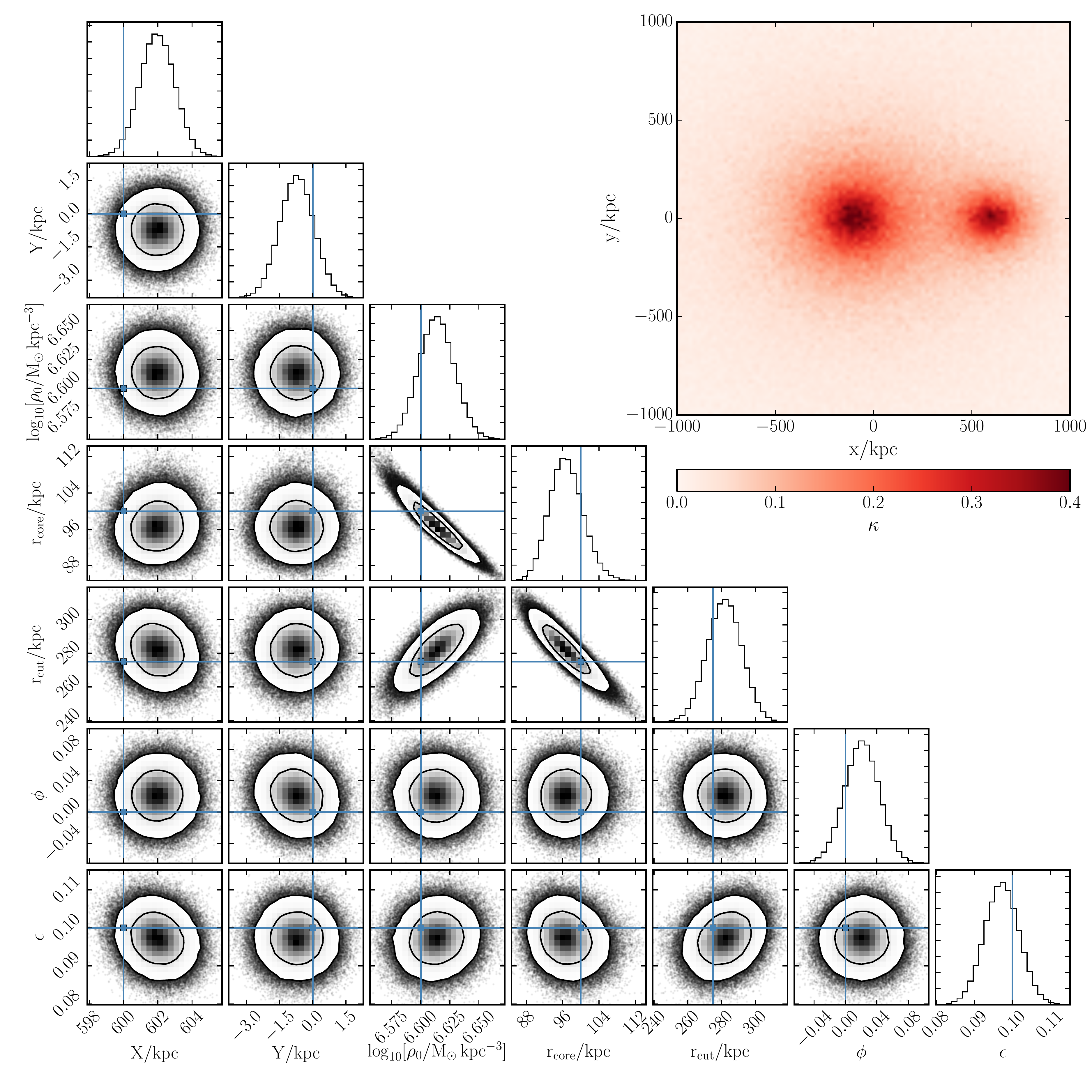}
	\caption{The posterior distributions for the model parameters of the smaller halo (on the right in the convergence map), found from simultaneously fitting two PIEMDs to the projected density generated from two model PIEMDs. The contours show 68 and 95\% confidence intervals. The model values used to generate the projected density are shown by the blue lines, and are recovered within the posterior distributions returned by the fitting procedure. The model values for the larger halo were also recovered, but are not shown here for clarity. The plot was made using \sc{corner.py} \citep{ForemanMackey2016}.}
	\label{fig:synthetic_nonoise_MCMC_kappa}
\end{figure*}

In Fig.~\ref{fig:synthetic_nonoise_MCMC_kappa} we show the results of fitting two PIEMDs to a synthetic density map. The synthetic map was generated by taking the projected density profile of two PIEMDs, here chosen to have parameters similar to that of the Bullet Cluster at the time at which it is observed, and then drawing a number of particles in each bin from a Poisson distribution with mean equal to the number of particles expected from the analytic profiles, assuming a particle mass equal to that used in our fiducial simulations.

The map in the top right of Fig.~\ref{fig:synthetic_nonoise_MCMC_kappa} shows visually the level of noise associated with having a discrete set of particles and using $20 \kpc$ bins, while the main corner plot shows that the fitting procedure recovers the input model within the error contours of the 2D projected posterior distributions. As the likelihood function in equation~\eqref{L_kappa} is based upon Poisson statistics in each bin, the width of the posterior distributions shows the uncertainty in model parameters due to having a finite number of simulation particles. Of particular interest is the width of the posterior of the halo position along the collision axis ($X$), as it is the separation of different components along this axis that can be used to infer the DM cross-section. Using the particle mass used in our simulations the width of the $X$ posterior distribution is $\sim 2 \kpc$.

\subsection{Parametric fits to shear maps}

\subsubsection{Generating shear maps}
\label{sect:gen_shear}

Although the projected density is technically observable through size and flux magnifications \citep[as recently done in ][]{2016MNRAS.457..764D}, weak lensing is usually done using the gravitational shear field. While the intrinsic ellipticities of galaxies are typically larger than the ellipticity from gravitational shear, with a large number of lensed galaxies the projected mass distribution of the lensing object can be determined.

The magnification of sources is described by the convergence, $\kappa$, while the distortion to the shape of galaxies is described by the shear $(\gamma_1 , \gamma_2)$. Here $\gamma_1$ describes stretching and squashing along the $x$-axis, while $\gamma_2$ describes these at 45\degree to the $x$-axis. In fact, the effect of lensing on galaxy ellipticities is described by the reduced shear, $g = \gamma / (1-\kappa)$. The quantities $\kappa$, $\gamma_1$ and $\gamma_2$ can all be related to the effective lensing potential, $\Psi$, through
\begin{equation}
\kappa = \frac{1}{2} \left(\frac{\partial^2 \Psi}{\partial x^2} + \frac{\partial^2 \Psi}{\partial y^2} \right),
\label{kappa_psi}
\end{equation}
\begin{equation}
\gamma_1 = \frac{1}{2} \left(\frac{\partial^2 \Psi}{\partial x^2} - \frac{\partial^2 \Psi}{\partial y^2} \right),
\label{gamma1_psi}
\end{equation}
and
\begin{equation}
\gamma_2 = \frac{\partial^2 \Psi}{\partial x \partial y}.
\label{gamma2_psi}
\end{equation}

The convergence is also given by the scaled projected density
\begin{equation}
\kappa(x,y) = \frac{\Sigma(x,y)}{\Sigma_\mathrm{crit}},
\label{kappa_crit}
\end{equation}
where the critical surface density, $\Sigma_\mathrm{crit}$, is dependent on the geometry of the source, observer and lens through
\begin{equation}
\Sigma_\mathrm{crit} = \frac{c^2}{4 \pi G} \frac{D_\mathrm{s}}{D_\mathrm{l}D_\mathrm{ls}},
\label{sigma_crit}
\end{equation}
where $D_\mathrm{s}$, $D_\mathrm{l}$, and $D_\mathrm{ls}$ are the angular diameter distances between the observer and the source, observer and lens, and lens and source respectively.

Using equations \eqref{kappa_crit} and \eqref{sigma_crit}, we can generate a $\kappa$ map from a simulation snapshot by binning the 2D particle distribution, having projected along the third dimension. Using the number of particles in a bin, the particle mass, and the bin area, we can calculate a projected density, $\Sigma(x,y)$. Then given a redshift for the lens (here $z=0.296$ as for the observed bullet cluster) and a redshift for the source galaxies (here we use $z_\mathrm{s} = 1$) we can calculate the critical surface density, which for our choice of cosmology was $\Sigma_\mathrm{crit} = \num{2.85e9} \, \msun \kpc^{-2}$.

Once we have a $\kappa$ map, we can generate maps of $\gamma_1$ and $\gamma_2$ by making use of equations (\ref{kappa_psi} - \ref{gamma2_psi}). Taking the Fourier transform of these equations, we find
\begin{equation}
\hat{\kappa} = -\frac{1}{2} (k_x^2 + k_y^2) \hat{\Psi},
\label{FT_kappa_psi}
\end{equation}
\begin{equation}
\hat{\gamma_1} = -\frac{1}{2} (k_x^2 - k_y^2) \hat{\Psi},
\label{FT_gamma1_psi}
\end{equation}
and
\begin{equation}
\hat{\gamma_2} = - k_x k_y \hat{\Psi},
\label{FT_gamma2_psi}
\end{equation}
where $\boldsymbol{k} = (k_x, k_y)$ is the wave vector conjugate to $\boldsymbol{x} = (x, y)$. These can be rearranged to give
\begin{equation}
\hat{\gamma_1} = \frac{k_x^2 - k_y^2}{k_x^2 + k_y^2} \hat{\kappa},
\label{gamma1_kappa}
\end{equation}
and
\begin{equation}
\hat{\gamma_2} = \frac{2 k_x k_y}{k_x^2 + k_y^2} \hat{\kappa}.
\label{gamma2_kappa}
\end{equation}
Finding $\gamma_1$ and $\gamma_2$ is then simply a case of taking the Fourier transform of $\kappa$, multiplying by the appropriate function of $k_x$ and $k_y$ and taking the inverse Fourier transform to return the desired shear component. The two components of $g$ are then given by these shear components divided by $1-\kappa$.

\subsubsection{Shear map likelihood function}

Given maps of the two reduced shear components generated from a simulation snapshot, $g_{1,i}^\mathrm{d}$ and $g_{2,i}^\mathrm{d}$, we can calculate a likelihood function
\begin{equation}
\mathcal{L}(\theta) = \prod_{i} \exp{\left( \frac{(g_{1,i}^\mathrm{d} - g_{1,i}^\mathrm{m})^2}{2 \sigma_\gamma^2} \right)} \exp{\left( \frac{(g_{2,i}^\mathrm{d} - g_{2,i}^\mathrm{m})^2}{2 \sigma_\gamma^2} \right)},
\label{L_shear}
\end{equation}
where  $g_{1,i}^\mathrm{m}$ and  $g_{2,i}^\mathrm{m}$ are the maps generated from the parametric model described by $\theta$. When reconstructing a shear field from the ellipticities of lensed galaxies, the variance of each component of the shear field at a pixel, $\sigma_\gamma^2 = \sigma_\mathrm{int}^2 +  \sigma_\mathrm{meas}^2$, comes from the intrinsic ellipticities of galaxies as well as shape measurement errors. Shape measurement errors depend on the quality of the data, as well as the method used to measure shapes, while the intrinsic ellipticities of galaxies are an unavoidable limitation to lensing measurements using gravitational shear. We thus set $\sigma_\mathrm{meas} = 0$ in this work, and assume that the only limitation to reconstructing a mass model using weak lensing comes from the number density of galaxies and the width of their intrinsic ellipticity distribution. \citet{2007ApJS..172..219L} found that for each galaxy $\sigma_\mathrm{int} \sim 0.26$ across a wide range of sizes, magnitudes and redshifts. Thus, given a number of lensed galaxies, $N$, within a pixel of a shear map, the contribution of intrinsic ellipticities to the average ellipticity of galaxies in that bin will be normally distributed with zero mean and standard deviation $\sigma_\gamma = 0.26/\sqrt{N}$.

In this work we use a square shear map with a side length of 3 Mpc, centred on the centre of mass of the two haloes. We first produce a convergence map of this same area, and then generate a shear map from this following the procedure described in \S\ref{sect:gen_shear}. In order to avoid wraparound errors, the convergence map is zero-padded up to a side length of 10 Mpc. The posterior distribution for parameters describing two elliptical PIEMDs can then be calculated as for the projected density, using the likelihood function in equation~\eqref{L_shear}, where $\sigma_\gamma$ is calculated assuming a source-galaxy density of 80 galaxies arcmin$^{-2}$. We also mask out any pixels where $\kappa > 0.6$, as in these regions the reduced shear can become very large and then individual pixels dominate the likelihood, these regions are approaching or in the strong lensing regime, and would not typically enter a weak lensing analysis.

\begin{figure*}
        \centering
        \includegraphics[width=\textwidth]{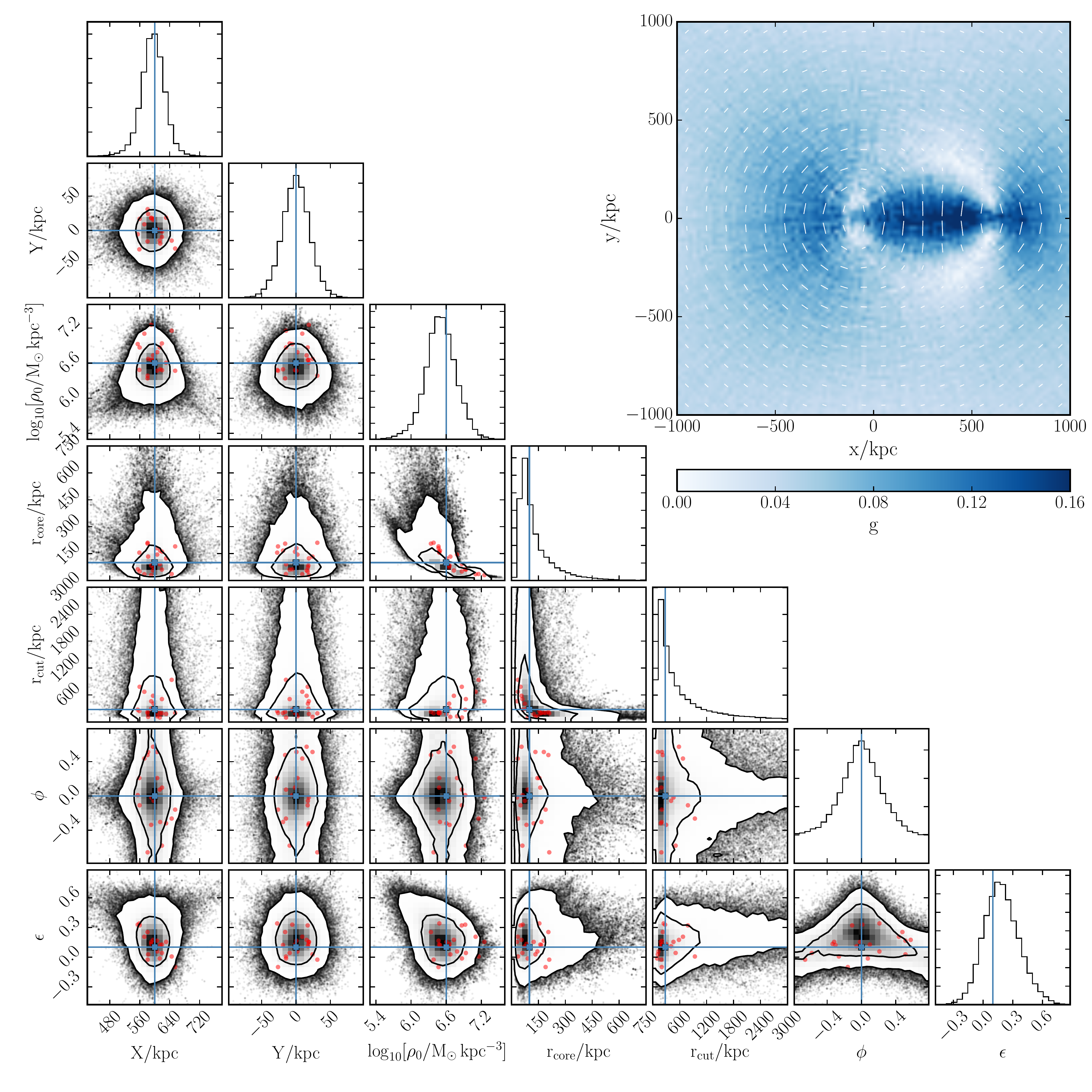}
	\caption{Similar to Fig.~\ref{fig:synthetic_nonoise_MCMC_kappa}, but for the case of fitting to reduced gravitational shear. The red points indicate the maximum likelihood parameter values found from fitting to the underlying shear map from the model with the addition of 20 different realisations of noise from source-galaxy intrinsic ellipticities. In the shear map in the top-right the colour represents the value of the reduced shear, while the white lines show the direction.}
	\label{fig:synthetic_particlenoise_MCMC_g}
\end{figure*}

The result of fitting to a shear map generated from the projected density profile in Fig.~\ref{fig:synthetic_nonoise_MCMC_kappa} is shown in Fig.~\ref{fig:synthetic_particlenoise_MCMC_g}. Unlike the case of fitting to the projected density, the width of the posterior distribution is no longer driven by the number of simulation particles, but by our greater  uncertainty on the shear map from the intrinsic shapes of lensed galaxies. The synthetic shear map generated (and shown in the top-right of Fig.~\ref{fig:synthetic_particlenoise_MCMC_g}) did not include any shape noise, and so the posterior distributions returned are centred on the true model values. The width of the posterior describes the range of results one would expect to derive had there been shape noise, as demonstrated by the red dots which show the maximum likelihood parameter values for 20 different realisations of maps where Gaussian noise was added to the synthetic shear map, with the variance of the noise corresponding to $\sigma_\gamma^2$.

The width of the posterior distributions in Fig.~\ref{fig:synthetic_particlenoise_MCMC_g} suggest that using gravitational shear with 80 galaxies arcmin$^{-2}$ we cannot determine the position of the bullet DM halo to better than $\sim \pm 40 \kpc$. This is consistent with \citet{2015Sci...347.1462H} who found a typical 1$\sigma$ error of $60 \kpc$ on the DM halo positions determined from weak gravitational lensing with $\sim 60$ galaxies arcmin$^{-2}$ (D. Harvey, private communication).

\section{Results}
\label{sect:results}

\subsection{Offsets with different cross-sections}

\begin{figure}
        \centering
        \includegraphics[width=\columnwidth]{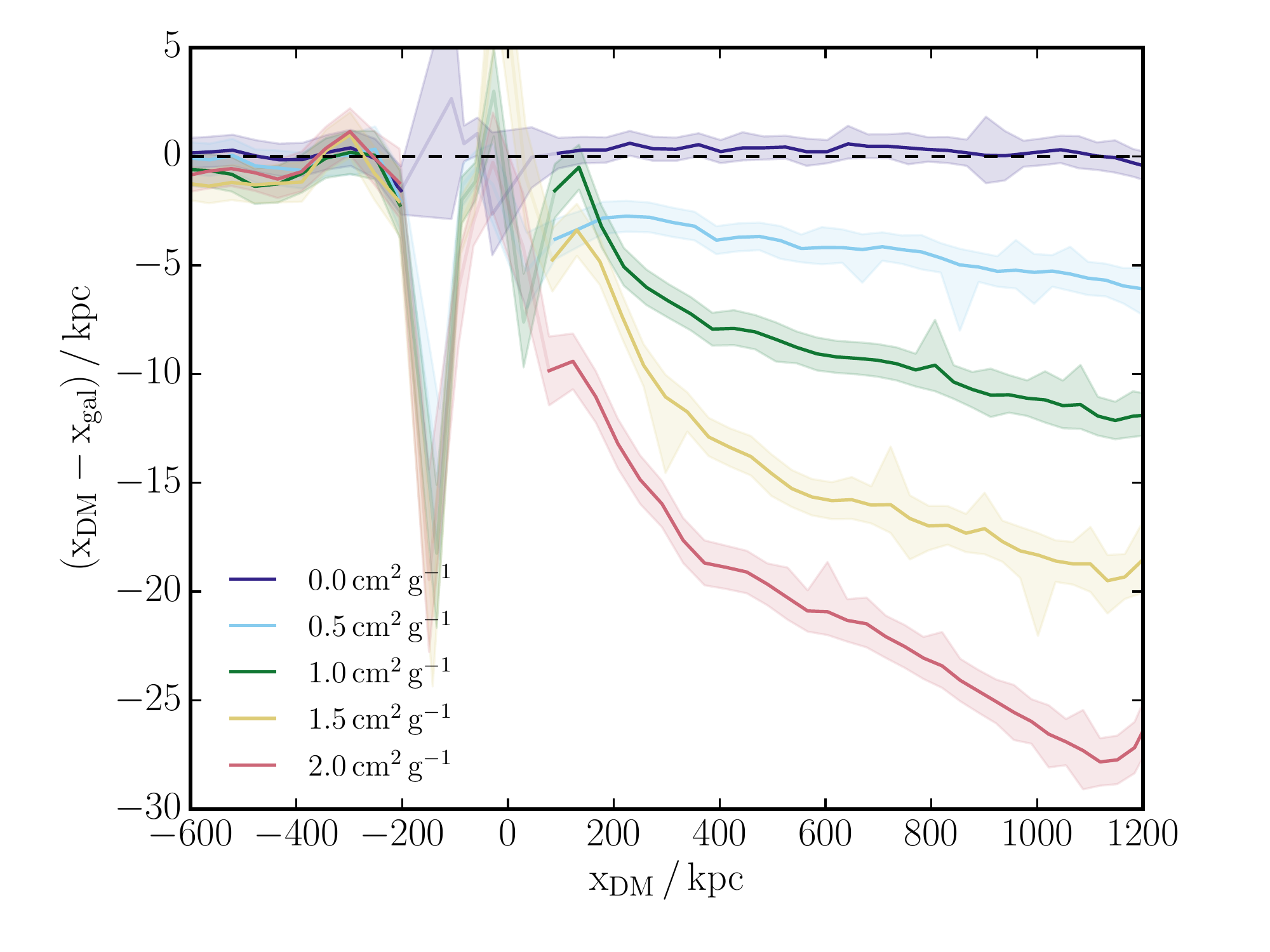}
	\caption{The offset between the stellar (galaxy) and DM component of the bullet halo for different SIDM cross-sections, where both the stellar and DM positions were determined by simultaneously fitting two PIEMDs to the respective projected maps. The offsets scale linearly with DM cross-section, and at the time of the observed bullet cluster the DM trails the galaxies by $\sim 10 \kpc$ for $\sigma / m = 1 \cmsg$. Lines are semi-transparent around the time of core-passage (which due to tidal forces happens at $X_\mathrm{DM} \approx -80 \kpc$ before the centres of mass of the two haloes meet) due to a degeneracy in the positions of the two haloes leading to spurious offsets.}
	\label{fig:offsets_nogas_ePIEMD}
\end{figure}

As our fiducial method to measure the positions of both stars and DM, we fit two PIEMDs to the projected surface density as described in \S\ref{sect:fit_to_kappa}. Doing this independently for the DM and stellar component we can then measure the offset along the collision axis between the two components. This was done for collisionless DM as SIDM with four different cross-sections. The offset between the stars and DM of the smaller bullet halo is shown as a function of the position of this halo in Fig.~\ref{fig:offsets_nogas_ePIEMD}. This position was measured along the collision ($x$) axis, relative to the centre of mass of the two haloes. As the main halo is substantially more massive than the bullet halo, this position is similar to the separation between the two DM haloes. For collisionless DM, the observed DM halo separation of $720 \kpc$ occurs when the bullet halo is at $X_\mathrm{DM} \approx 600 \kpc$.

The offsets scale linearly with cross-section, in agreement with R08 and K14, but the size of the offsets for a given cross-section are considerably smaller than those found in R08, and about 40\% smaller than in K14. For $\sigma / m = 1 \cmsg$ the offset at the time of the observed Bullet Cluster is $\sim 10 \kpc$, whereas R08 find that a similar cross-section leads to the DM trailing the galaxies by almost 40 kpc. From the observed trailing of galaxies by DM of $25 \pm 29 \kpc$ R08 placed constraints on the DM cross-section of $\sigma / m < 1.25 \cmsg$, whereas all of our simulated cross-sections would be consistent with this observation. This discrepancy is investigated in the following two sections, where we vary our initial conditions, and then the method used to measure positions.

\subsection{Sensitivity to varying initial conditions}
\label{sect:varICs}

The offsets for different cross-sections depend on the initial conditions used, as changing the masses and concentrations of the haloes changes the rate of DM scattering as well as the gravitational forces that dominate the dynamics of the merger. In this section, we investigate changing the initial conditions. We vary one parameter at a time from its value in our fiducial model, and change the parameters in a way that has been used in previous simulations or has been hinted at by previous results.

\subsubsection{Main halo concentration}

The first parameter we vary is the concentration of the main halo. Our fiducial model used $c=3$ as this was found to be required by SF07 to reproduce the observed offset between the bright X-ray bullet and the associated DM halo. This result used the rather limiting assumption (as used in this work) that the gas density initially follows the DM density. \citet{2014ApJ...787..144L} used more complicated models for their initial conditions, with triaxial DM haloes, and a seven parameter model for the gas profile of each halo. They found their best fitting model to have $c = 1.17 \pm 0.14$ for the main halo, which would put this halo well below the median concentration-mass relation. To investigate how a low initial concentration for the main halo affects our results, we ran simulations with an initial concentration for the main halo of $c=1.94$, which was the best fit concentration for the main halo measured after the collision via weak lensing. 

\begin{figure}
        \centering
        \includegraphics[width=\columnwidth]{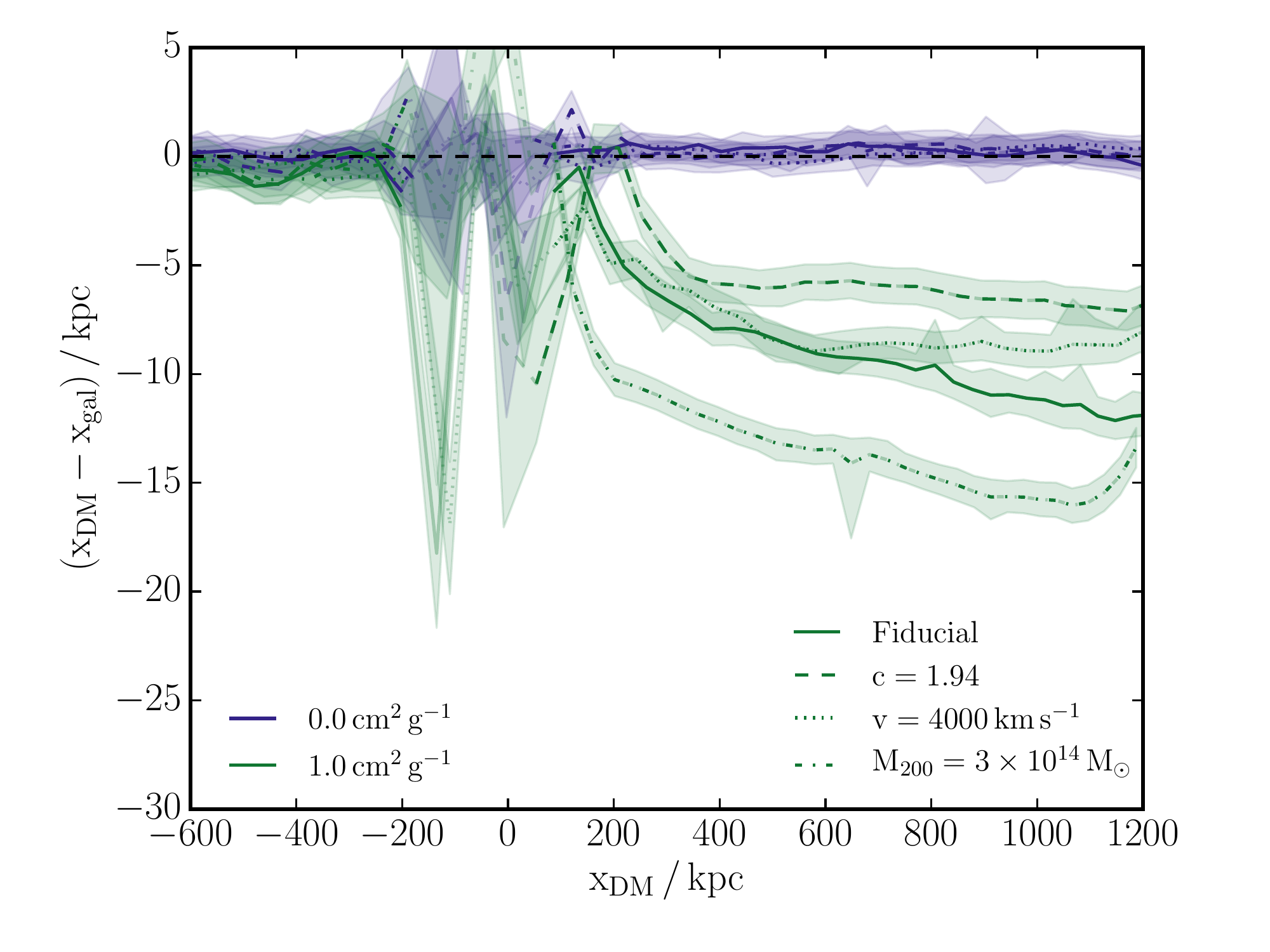}
	\caption{DM--galaxy offsets as in Fig.~\ref{fig:offsets_nogas_ePIEMD}, but with four different sets of initial conditions, each run with collisionless DM and SIDM with $\sigma / m = 1 \cmsg$. These initial conditions are described in \S\ref{sect:varICs}, but in summary are as follows: compared to our fiducial model `$c=1.94$' has a lower concentration for the main halo, `$v = 4000 \kms$' has an increased relative velocity between the two haloes, and `$M_{200} = \num{3e14} \msun$' has a more massive bullet halo.}
	\label{fig:offsets_altIC}
\end{figure}

The resulting offsets with collisionless DM and SIDM with $\sigma / m = 1 \cmsg$ are shown in Fig.~\ref{fig:offsets_altIC}. The offset with SIDM is reduced relative to the fiducial model, which is to be expected given that with a lower concentration, the projected density through the centre of the halo is reduced. This means that particles in the bullet halo, which has zero impact parameter and passes through the centre of the main halo, pass through less DM and are less likely to scatter from particles in the main halo. In fact, the fraction of DM particles from the bullet halo that scatter with particles from the main halo drops from 13\% for our fiducial model to 10\%, in broad agreement with estimates of the scattering fraction that can be made from Fig.~\ref{fig:projected_density_comparison}.


\subsubsection{Relative velocity between haloes}

As discussed in \S\ref{sect:relative_vel} the shock velocity in the observed Bullet Cluster is $4700 \pm 600 \kms$. In previous work using the Bullet Cluster to constrain SIDM (R08, K14) this has been used as the relative velocity between the two DM haloes, despite hydrodynamical simulations showing that the relative velocity of the shock front and pre-shocked gas in Bullet Cluster-like simulations is significantly higher than the relative velocity of the DM haloes \citep[SF07,][]{Milosavljevic:2007hf,2014ApJ...787..144L}.

In Fig.~\ref{fig:offsets_altIC} we show how the offset between the bullet DM halo and galaxies changes when the collision velocity is increased. We start the haloes with a relative velocity of $4000 \kms$ at a separation of $4 \mpc$, which leads to a relative velocity of $4700 \kms$ at the time of the observed Bullet Cluster. This is in contrast with our fiducial model, where haloes start with the velocity corresponding to falling from rest at infinite distance, and the relative velocity between DM haloes is $3900 \kms$ at the time of the observed Bullet Cluster. We find that the offsets are not very sensitive to this change in relative velocity.

\subsubsection{Mass of bullet halo}

The weak lensing derived mass for the bullet halo of $M_{200} = \num{1.5e14} \msun$ is low in comparison with the strong lensing results \citep{2006ApJ...652..937B} that calculate the mass in a $300 \kpc$ radius cylinder centred on the bullet halo's BCG to be $\num{3e14} \msun$. While this is the total mass in this region, and includes a contribution from the main halo, this is still suggestive that the weak lensing mass may be an underestimate. Simulations that have looked to reproduce the gas morphology and luminosity have also found best fit mass-ratios for the merger between 7:1 and 5:1 \citep{Milosavljevic:2007hf,2008MNRAS.389..967M,2014ApJ...787..144L}.

For these reasons we run simulations with an increased mass for the bullet halo of $M_{200} = \num{3e14} \msun$, keeping the concentration the same as in our fiducial model. This leads to a significant increase in the separation between DM and galaxies in the bullet, with the offset at the time of the observed Bullet Cluster increasing from $10 \kpc$ for our fiducial model to $14 \kpc$.

\subsubsection{Impact parameter}

While the gas morphology implies a collision that was close to head-on, the bright gas bullet is not located precisely along the line connecting the centres of the two cluster haloes, suggesting a small non-zero impact parameter. We therefore run simulations with off-centre collisions, and investigate how sensitive the DM-galaxy offsets are to this change.

We continue to start the simulations with the two haloes separated by $4 \mpc$ and on a zero energy orbit, but rotate the velocities of the haloes by $\theta_\mathrm{init}$ with respect to the $x$-axis that connects the two halo centres (keeping the velocities of the two haloes anti-parallel). We choose $\theta_\mathrm{init}$ such that the two haloes would have a closest approach of $r_*$ if they behaved as point masses throughout the merger. The force between the two haloes is reduced (compared to the case of point masses) when their mass distributions overlap, so that the actual minimum separation between the halo centres, $r_\mathrm{min}$, is significantly larger than $r_*$.

We summarise our different impact parameter runs in Table~\ref{table:rmin}. As well as $r_*$, $\theta_\mathrm{init}$, and $r_\mathrm{min}$, we include the perpendicular distance between the two haloes' velocities when they are separated by $4 \mpc$, $b_{4 \mpc}$, and the angle between the halo-halo separation and the bullet halo velocity at the time of the observed Bullet Cluster, $\theta_\mathrm{obs}$. Assuming that gas is stripped in the opposite direction to the direction of motion, $\theta_\mathrm{obs}$ should roughly correspond to the angle between the DM-gas separation in the bullet halo and the DM-DM separation between the two haloes.

\begin{table}
\centering
\caption{Summary of non-zero impact parameter simulations.}
\label{table:rmin}
\begin{tabularx}{\columnwidth}{XXXXX}
\toprule
$r_* / \kpc$ & $\theta_\mathrm{init} / ^\circ$ & $b_{4 \mpc} / \kpc$   & $r_\mathrm{min} / \kpc$ & $\theta_\mathrm{obs} / ^\circ$\\
\midrule
0    & 0     & 0   & 0 & 0     \\
12.5 & 3.2 & 224 & 102 & 6    \\
25   & 4.5 & 316 & 153  & 10  \\
50   & 6.4 & 447 & 236  & 18  \\
100  & 9.1 & 632 & 354 & 30  \\
\bottomrule
\end{tabularx}
\end{table}

\begin{figure}
        \centering
        \includegraphics[width=\columnwidth]{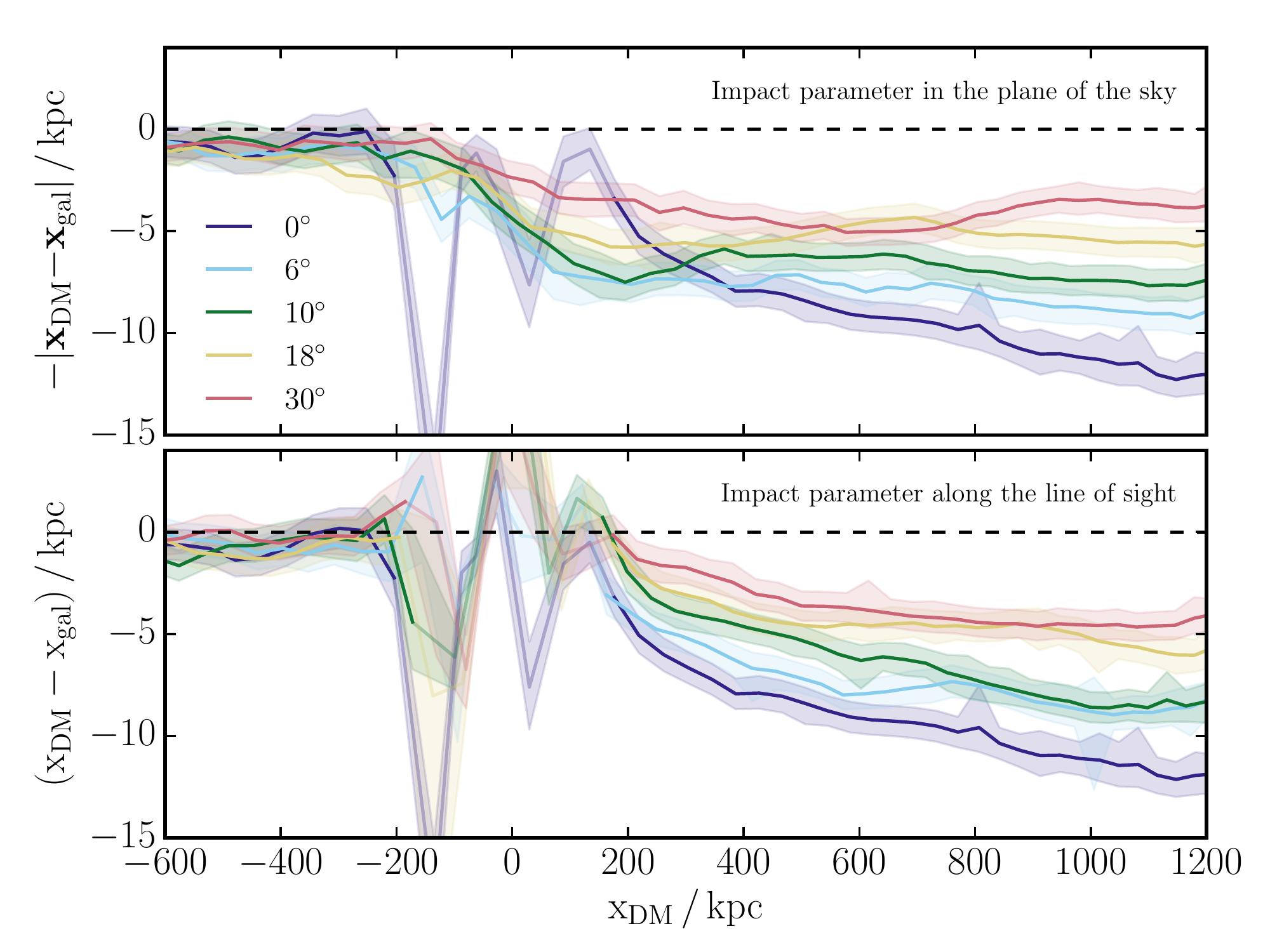}
	\caption{DM-galaxy offsets with $\sigma / m = 1 \cmsg$ and four different impact parameters, as well as a head-on collision. The runs are labelled by the angle between the separation of the two haloes and the velocity of the bullet halo measured at the time of the observed Bullet Cluster, $\theta_\mathrm{obs}$. The fitting was done by simultaneously fitting two PIEMDs to the projected mass distribution. In the top panel the impact parameter was in the plane of the sky, while in the bottom panel it was along the line of sight.}
	\label{fig:offsets_rmin}
\end{figure}

In Fig.~\ref{fig:offsets_rmin} we plot the separation between the DM and galaxies with different impact parameters. In the top panel the impact parameter is in the plane of the sky, while in the bottom panel it is along the line of sight, and the collision appears as if head on. Note that in the top panel we measure the 2D offset between the DM and galaxies, as this offset is no longer along the $x$-axis.

We find that moderate impact parameters only have a small effect on the DM-galaxy offsets. SF07 found that $r_* < 12.5 \kpc$ to avoid a gas distribution that is more asymmetric than that observed, while \citet{2008MNRAS.389..967M} found that an impact parameter, $b=150 \kpc$, gave the best match to the gas morphology and relative X-ray brightness of the two gaseous haloes. This means that even our smallest non-zero impact parameter is large compared to that used for the best-fitting results from other simulations of the Bullet Cluster, and so we expect any impact parameter consistent with the observed Bullet Cluster to decrease the DM-galaxy offset by less than 20\%.

\subsection{Offsets with different position measures}
\label{sect:offsets_methods}

\begin{figure}
        \centering
        \includegraphics[width=\columnwidth]{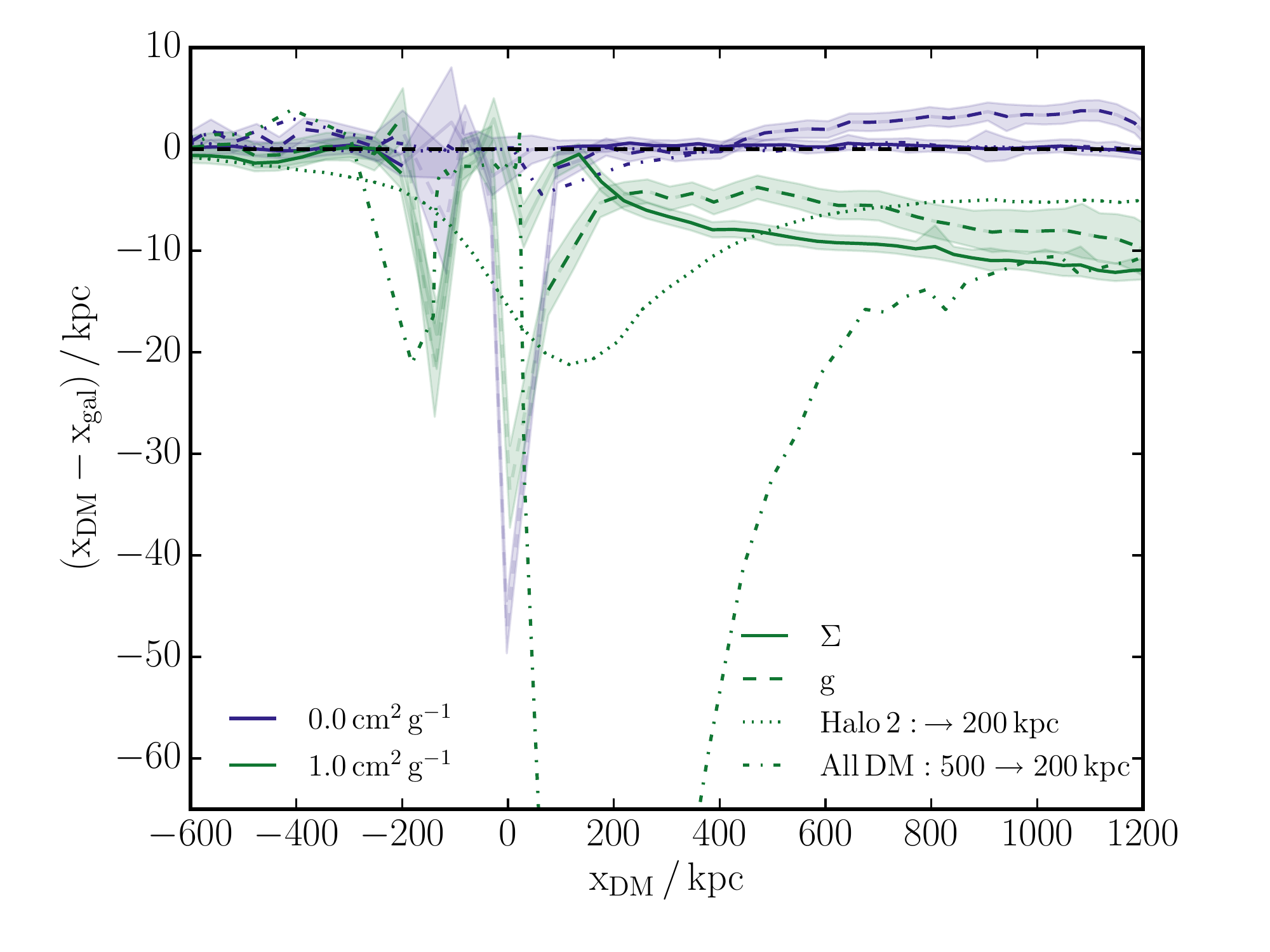}
	\caption{DM - galaxy offsets as in Fig.~\ref{fig:offsets_nogas_ePIEMD}, but measured using different methods: fitting to the projected surface density ($\Sigma$), the reduced gravitational shear ($g$) and two different shrinking circles techniques. For all methods but $g$, the same method was used to find the position of both the stars and the DM, while for $g$ it was only the DM measured using shear with the stars being measured using $\Sigma$. The two shrinking circles techniques are described in \S\ref{sect:offsets_methods}.}
	\label{fig:offsets_nogas_methods}
\end{figure}

Having found that the offset results are reasonably insensitive to the choice of initial conditions, in Fig.~\ref{fig:offsets_nogas_methods} we show the effect of using different methods to measure positions. For both collisionless DM and SIDM with $\sigma / m = 1 \cmsg$ we measured the separation between the stellar and DM components of the bullet halo using the methods described in \S\ref{sect:measuring_positions}. For the shrinking circles and projected density measurements the same method was used for finding the position of both the stars and DM, while for the shear measurement the separation is that between the stellar halo measured by fitting to the projected density and the DM halo measured using reduced gravitational shear. Fig.~\ref{fig:synthetic_particlenoise_MCMC_g} demonstrates that with 80 galaxies/arcmin$^2$ the position of the bullet halo can only be determined to $\pm 40 \kpc$. As this uncertainty is larger than the offsets for any of our simulated cross-sections, detecting SIDM using weak lensing and the Bullet Cluster alone would not be possible. In Fig.~\ref{fig:offsets_nogas_methods} the lines derived from reduced gravitational shear used 8000 galaxies/arcmin$^2$, giving errors indicative of what could be achieved with $\sim 100$ Bullet Cluster-like systems.   

As discussed in \S\ref{sect:shrinking_circles} the shrinking circles procedure will only return one position for a distribution of particles. We therefore use two different approaches to return the position of the bullet halo, both of which shrink the circle down to a final size of $R_\mathrm{min} = 200 \kpc$ as used by R08. The first method (\emph{Halo 2}) is to apply the shrinking circles procedure to only the particles that were originally part of the bullet halo. The second method (\emph{All DM}) is to apply the shrinking circles procedure to all of the DM, but starting with a circle centred on the second halo, as determined by Halo 2, with a starting radius of $500 \kpc$.

The different methods for measuring positions give very different results for the same SIDM cross-section, highlighting the importance of matching the analysis to what is done observationally. The offsets measured for $\sigma / m = 1 \cmsg$ using different methods can be as different as the offsets for the different cross-sections shown in Fig.~\ref{fig:offsets_nogas_ePIEMD}, particularly soon after core passage. Of particular note is the large offsets measured using shrinking circles on all of the DM. This method was also highly sensitive to the choice of starting position and starting radius, suggesting it is not a robust way to measure offsets from simulations. As a method similar to this was used by R08, this explains the large offsets and tight constraints on the DM cross-section that they found.

The sensitivity to the method used to measure the positions can be understood when one considers that after core passage of the two haloes, there are three distinct sets of DM particles: those originally from the main halo that have not interacted with any particles from the sub halo, those originally from the sub halo that have not interacted with any particles from the main halo, and particles from one halo that have scattered with a particle from the other halo.\footnote{For particles that are involved in an inter-halo scattering event, particles from the two haloes are indistinguishable when the scattering cross-section is isotropic.} The momentum transfer between the two haloes caused by isotropic DM elastic scattering acts differently to the stripping of gas due to hydrodynamical forces, as only a subset of DM particles receive a momentum kick. These particles then lag behind the halo from which they came, gravitationally pulling it back, but they do this equally to un-scattered DM particles and galaxies, and so do not lead to an offset between unscattered DM particles and the collisionless galaxies. Any offset found between the DM and galaxies is a result of fitting the wake of scattered particles and so depends sensitively on how positions are measured. 

\begin{figure}
        \centering
        \includegraphics[width=\columnwidth]{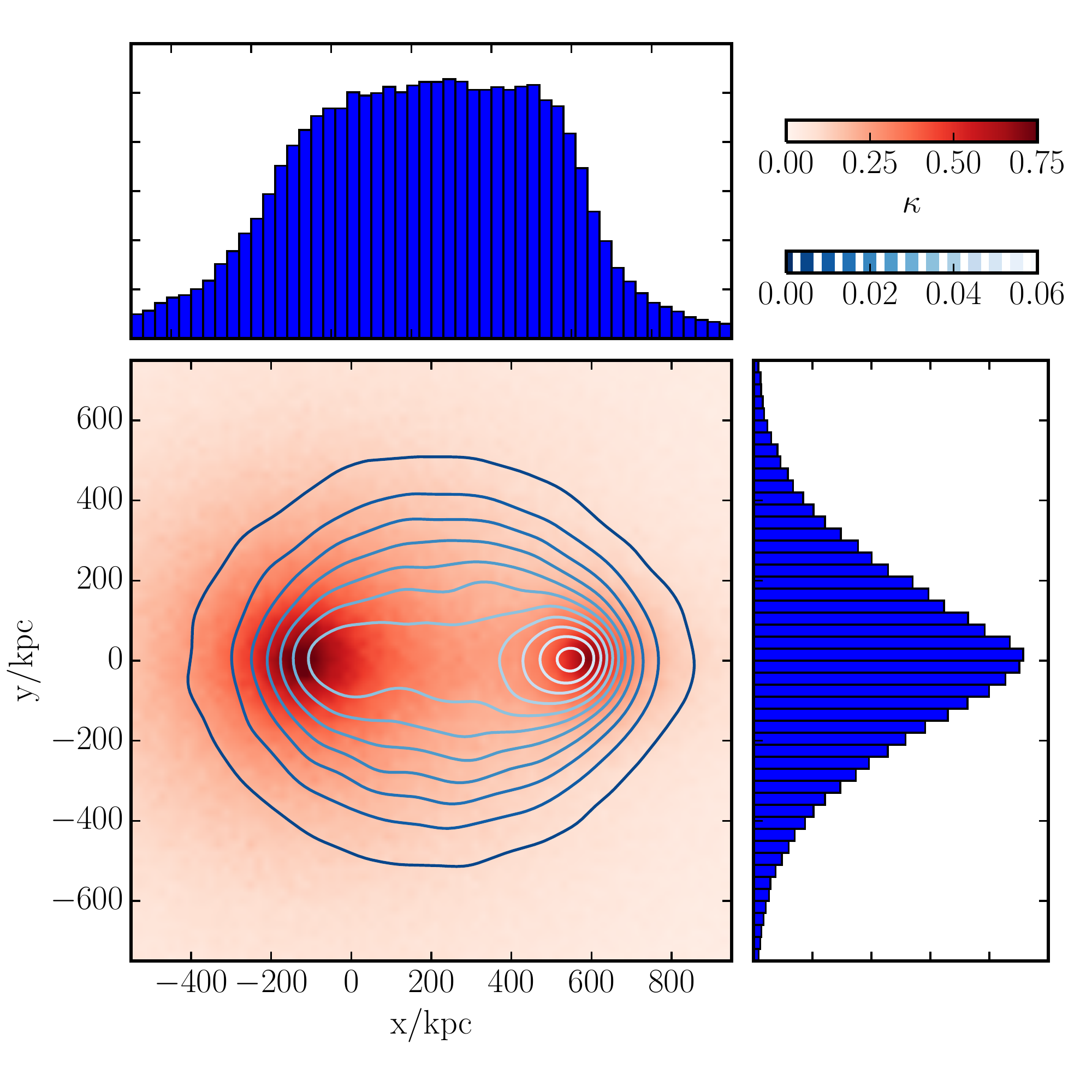}
	\caption{For SIDM with $\sigma / m = 1 \cmsg$ the projected density of all DM in red, with the projected density of DM particles that have scattered with a particle originally from the other DM halo shown in blue contours, and also projected along the axes and shown as 1D histograms. The total mass in these scattered particles is $\num{6.5e13} \msun$, corresponding to 13\% of particles from the bullet halo scattering with particles in the main halo.}
	\label{fig:scattered_distribution}
\end{figure}

For $\sigma / m = 1 \cmsg$ we show the projected DM density at the time of the observed Bullet Cluster in Fig.~\ref{fig:scattered_distribution}, along with the distribution of particles that have scattered with a particle from the other halo. For this cross-section and our fiducial initial conditions, 13\% of particles from the bullet halo scatter with particles from the main halo. The distribution of these scattered particles is quite broad, with the highest projected density of scattered particles being only 10\% of the total projected density at the same location.

\subsubsection{The position returned by shrinking circles to different final radii}
\label{App:shrinking_circles}

\begin{figure}
        \centering
        \includegraphics[width=\columnwidth]{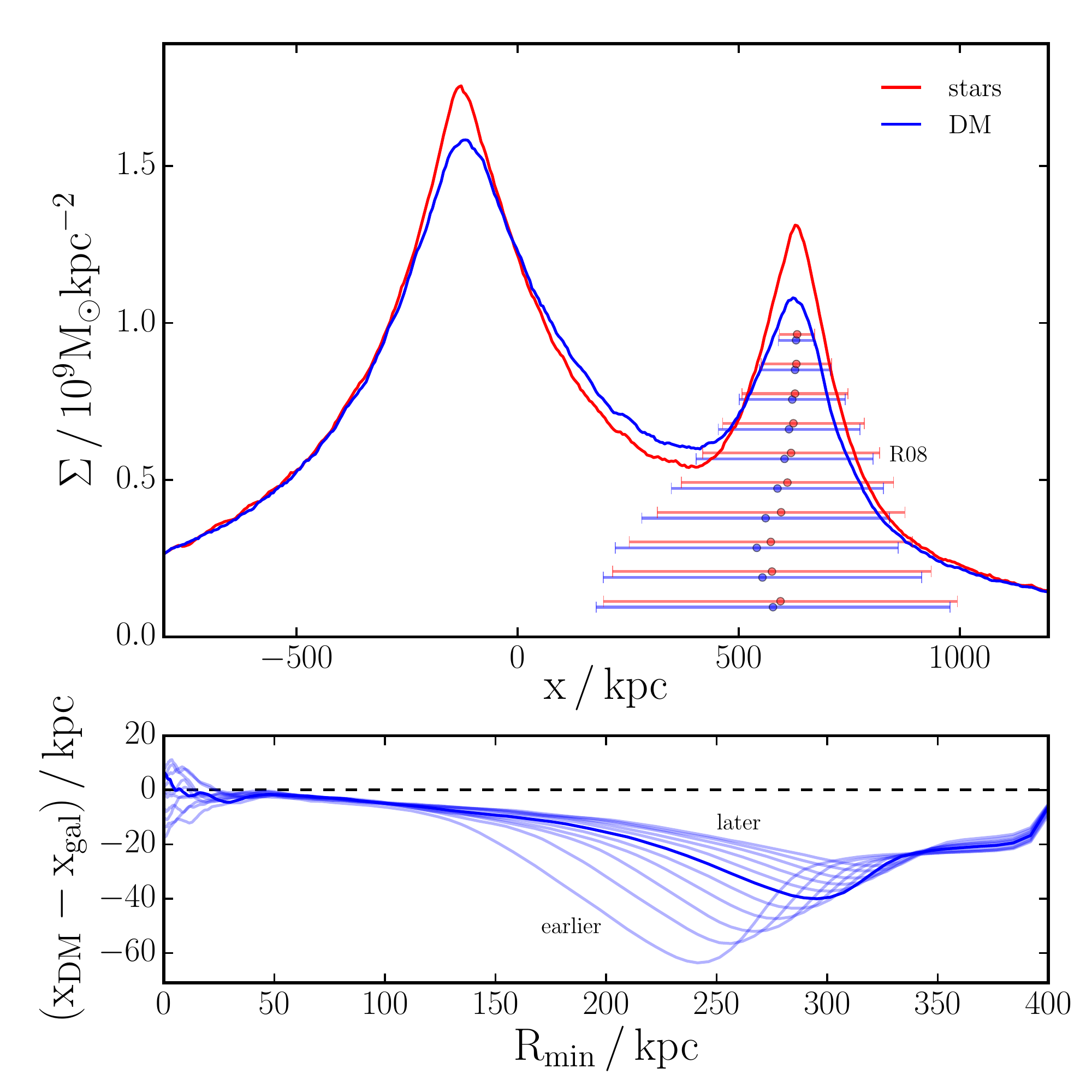}
	\caption{Top panel: the average projected surface density measured in a $400 \kpc$ strip centred on the collision axis, for both the stars and DM (the stellar surface densities have been scaled up so that the mass in stars matches that in DM). The points show the position of the bullet halo returned by the shrinking circles procedure with different $R_\mathrm{min}$, with the width of horizontal bars being twice $R_\mathrm{min}$. Bottom panel: the DM-galaxy offset as a function of $R_\mathrm{min}$. The dark line corresponds to the top panel (when the two haloes are separated by $\sim 720 \kpc$, and with $\sigma / m = 1 \cmsg$), while the lighter lines are for successive snapshots separated by 10 Myr.}
	\label{fig:shrinkcirc_X_R}
\end{figure}

To illustrate the problems with using a shrinking circles procedure to measure the positions of stars and DM we show an example in Fig.~\ref{fig:shrinkcirc_X_R}, run on our fiducial simulation with $\sigma / m = 1 \cmsg$ at the time of the observed Bullet Cluster (the same snapshot used for Fig.~\ref{fig:scattered_distribution}). The position returned for both the stars and DM varies as a function of $R_\mathrm{min}$, with the offset between the stars and DM also depending sensitively on $R_\mathrm{min}$.

An initial position for each of the stellar and DM components of the bullet halo is made by running shrinking circles on only particles that were part of the bullet halo in the initial conditions. The initial radius used was $400 \kpc$, a bit over half of the separation between the two DM haloes. Initially as the circles are shrunk and re-centred they shift left due to the gradient in density coming from the main halo. As this gradient is steeper closer to the main halo, the DM position (that initially lies to the left of the stellar position) is affected more by the presence of the main halo, which in turn leads to spuriously large offsets. As the circles are shrunk further, they centre in on a region dominated by the bullet halo, and the offsets decrease. Shrinking down to   $R_\mathrm{min} \lesssim 50 \kpc$ the results become noisy as the number of particles involved in the position estimate decreases, and there is no clear density peak (with $\sigma / m = 1 \cmsg$ the core size of the bullet halo is $\sim 100 \kpc$, though this is less obvious in the top of Fig.~\ref{fig:shrinkcirc_X_R} due to projection effects).

Even before the results become noisy, the offsets between the stellar and DM peaks become very small, in agreement with K14 who found that the peaks in stellar and DM projected density were perfectly coincident when DM scattering was isotropic. This raises the question of whether any constraints can be placed on \emph{isotropic} SIDM from looking at separations between local galaxy and DM peaks in colliding clusters. That being said, most studies that look for offsets between galaxies and peaks in free form lensing reconstructions either bin lensed galaxies \citep{2012ApJ...744...94R,2016MNRAS.459..517K} effectively smoothing the DM distribution on some scale, or use a regularisation scheme \citep[e.g.][]{2006ApJ...652..937B}, such that the diffuse cloud of scattered particles (Fig.~\ref{fig:scattered_distribution}) could shift the derived DM peak back and lead to a measurable offset.

From the bottom panel of Fig.~\ref{fig:shrinkcirc_X_R} it is clear that $R_\mathrm{min} = 200 \kpc$ can give misleadingly large offsets, which explains the tight constraints on the DM cross-section found by R08. What is also clear is that there is no good choice for $R_\mathrm{min}$, as the results do not converge as $R_\mathrm{min}$ is decreased. For these reasons we fit parametric models to our haloes in this paper, as is often done in gravitational lensing analyses \citep{2005MNRAS.359..417S,2010MNRAS.402L..44R,2012ApJ...757....2G,2015Sci...347.1462H,2015MNRAS.449.3393M,2016ApJ...820...43S}. While this does not directly relate to what was done in \citet{2006ApJ...652..937B}, where strong and weak lensing were combined to produce a non-parametric mass model of the Bullet Cluster, a mock strong-lensing analysis is beyond the scope of this paper. We cannot do strong lensing with our simulations as the surface density of our simulated bullet halo does not exceed the critical surface density for a lens at the Bullet Cluster's redshift. The absence of strong lensing with SIDM was noted by \citet{2001MNRAS.325..435M}, who found that with moderate cross-sections of $0.1 \-- 1 \cmsg$, the number of radial and giant-tangential arcs would fall well below what is observed. However, they point out that even with a collisionless DM simulation the number of strong lensing features falls below what is observed, and that bright central galaxies probably play an important role in generating strong lensing features. While this is certainly an interesting avenue to constrain SIDM, without including the effects of galaxy formation physics in our simulations, and with these simulations starting from idealised initial conditions, our work is not suited to testing whether the presence of strong lensing features can constrain the DM cross-section.

\subsection{Offsets including gas}
\label{sect:including_gas}

So far, the results have been from simulations without any gas. However, real galaxy clusters have significant gas fractions. While there is less gas than DM, the additional hydrodynamic forces that act on the gas can alter the dynamics of merging clusters. In this section we look at the changes from the previous results when each halo contains an adiabatic gas component making up 16\% of the total halo mass.

\begin{figure}
        \centering
        \includegraphics[width=\columnwidth]{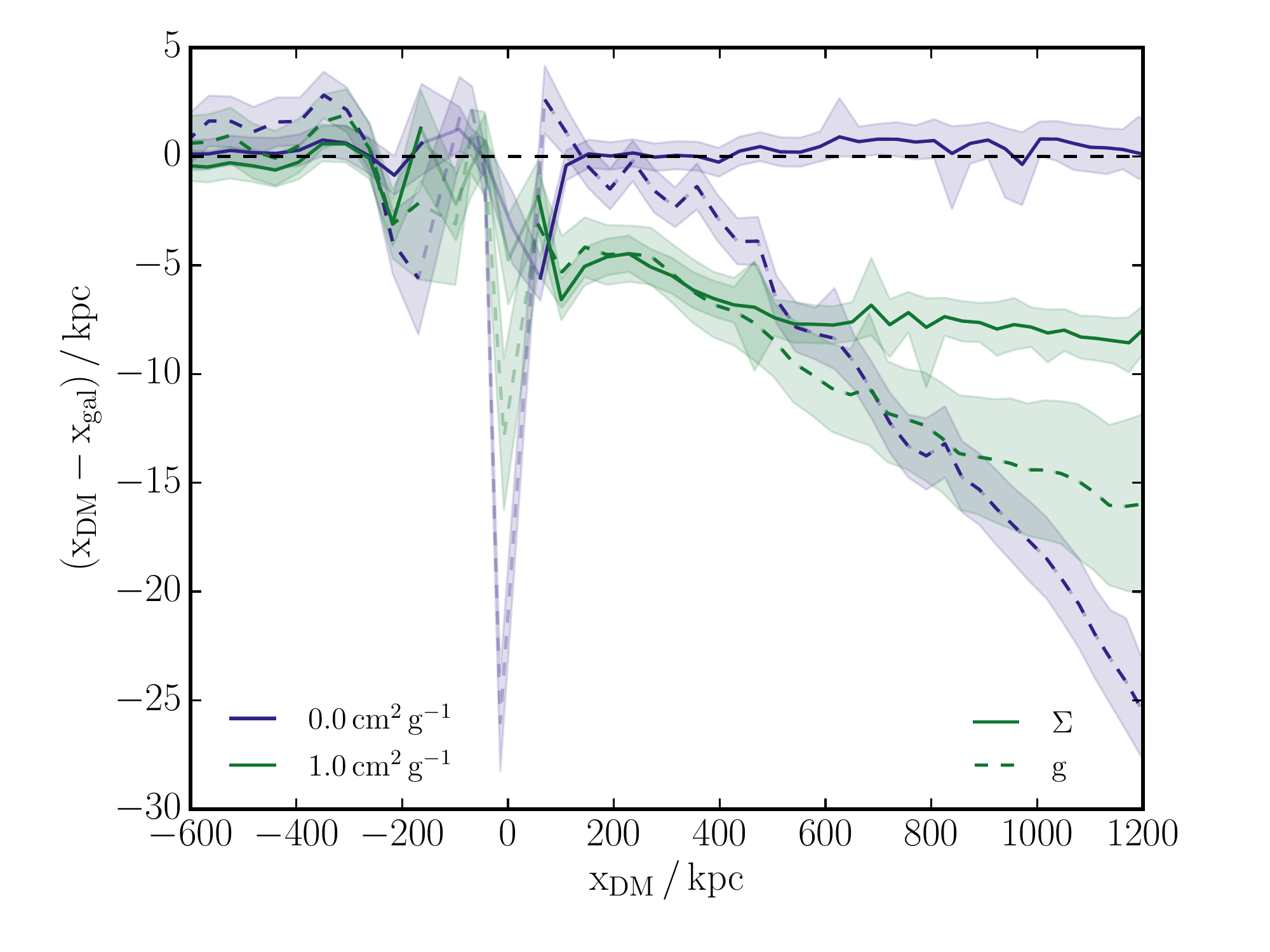}
	\caption{The offset between the stars and DM from simulations including adiabatic gas. As in Fig.~\ref{fig:offsets_nogas_methods}, the $g$ measurement is the offset between the DM position measured using reduced shear and the stellar position measured by fitting to the projected density of stars. As such, the large offset with collisionless DM which is not seen when both the stellar and DM positions are measured from their projected densities, means that fitting to the projected density or reduced shear of the same mass distribution, can lead to strongly differing results. }
	\label{fig:separations_M200match_gas}
\end{figure}

The resulting offsets between stars and DM are shown in Fig.~\ref{fig:separations_M200match_gas}. The offsets measured for $\sigma / m = 1 \cmsg$ remain largely unchanged, with a small decrease (compared with the gas-free case) in the offset measured by fitting to the projected surface density. This results from the decreased optical depth for scattering as particles pass through the main halo, owing to $\sim 16\%$ of the DM mass now being in the form of gas. Most strikingly, there is now a significant offset measured with collisionless DM when measuring the DM position using gravitational shear. This is surprising, particularly as our DM and stars have the same phase space distribution at all times in our collisionless DM run, so this offset is a result of different fitting methods returning substantially different position estimates.

\subsubsection{Explaining offsets with collisionless DM}

\begin{figure}
        \centering
        \includegraphics[width=\columnwidth]{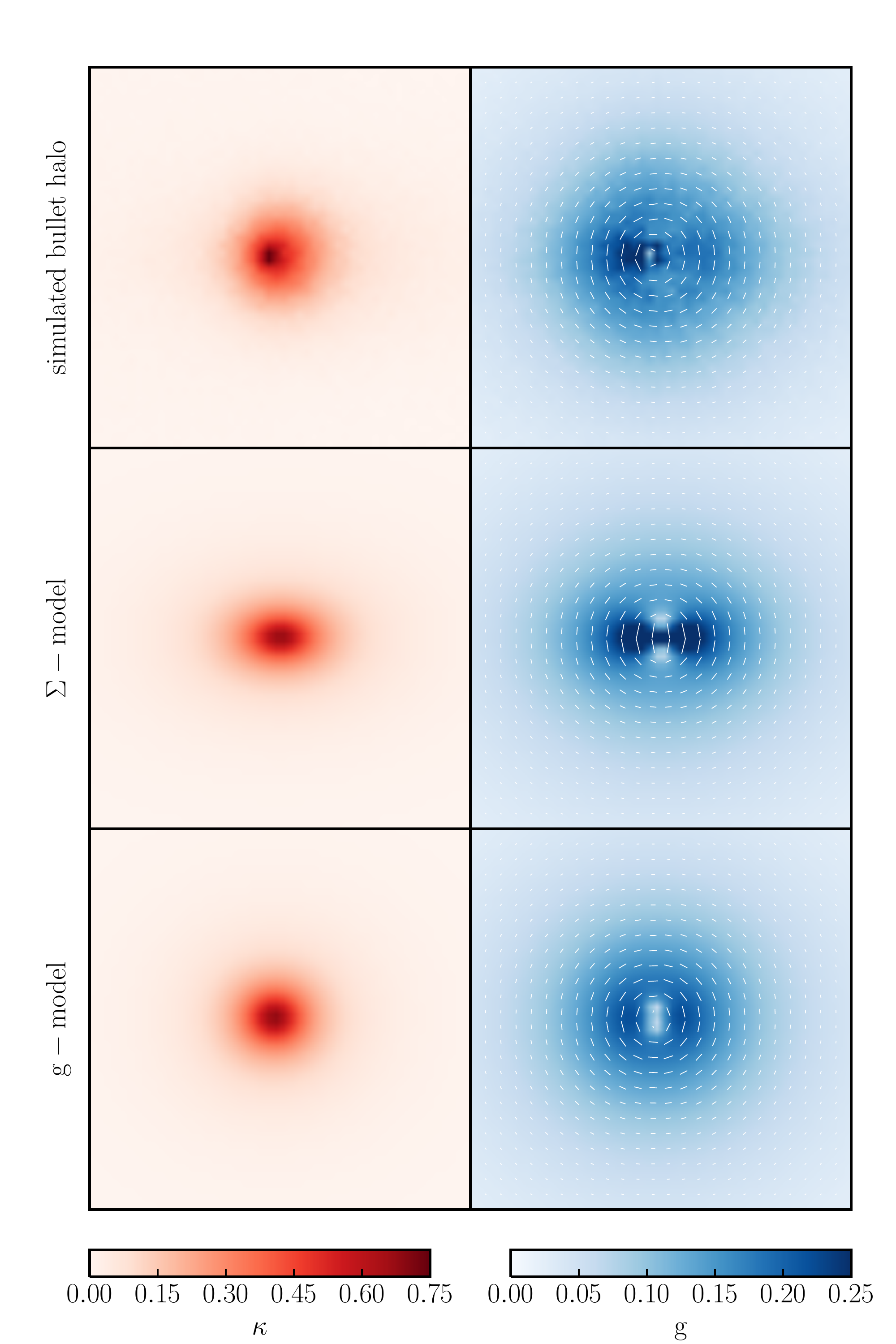}
	\caption{The convergence (left column) and reduced shear (right column) due to the bullet halo DM, for a simulation with collisionless DM and non-radiative gas. The top row shows the simulation output (only including DM particles that are part of the bullet halo in the initial conditions), while the middle and bottom rows show the best fit maps generated by fitting to the projected surface density and reduced gravitational shear respectively. Each panel is $1\,$Mpc across.}
	\label{fig:map_of_halo2}
\end{figure}

\begin{figure}
        \centering
        \includegraphics[width=\columnwidth]{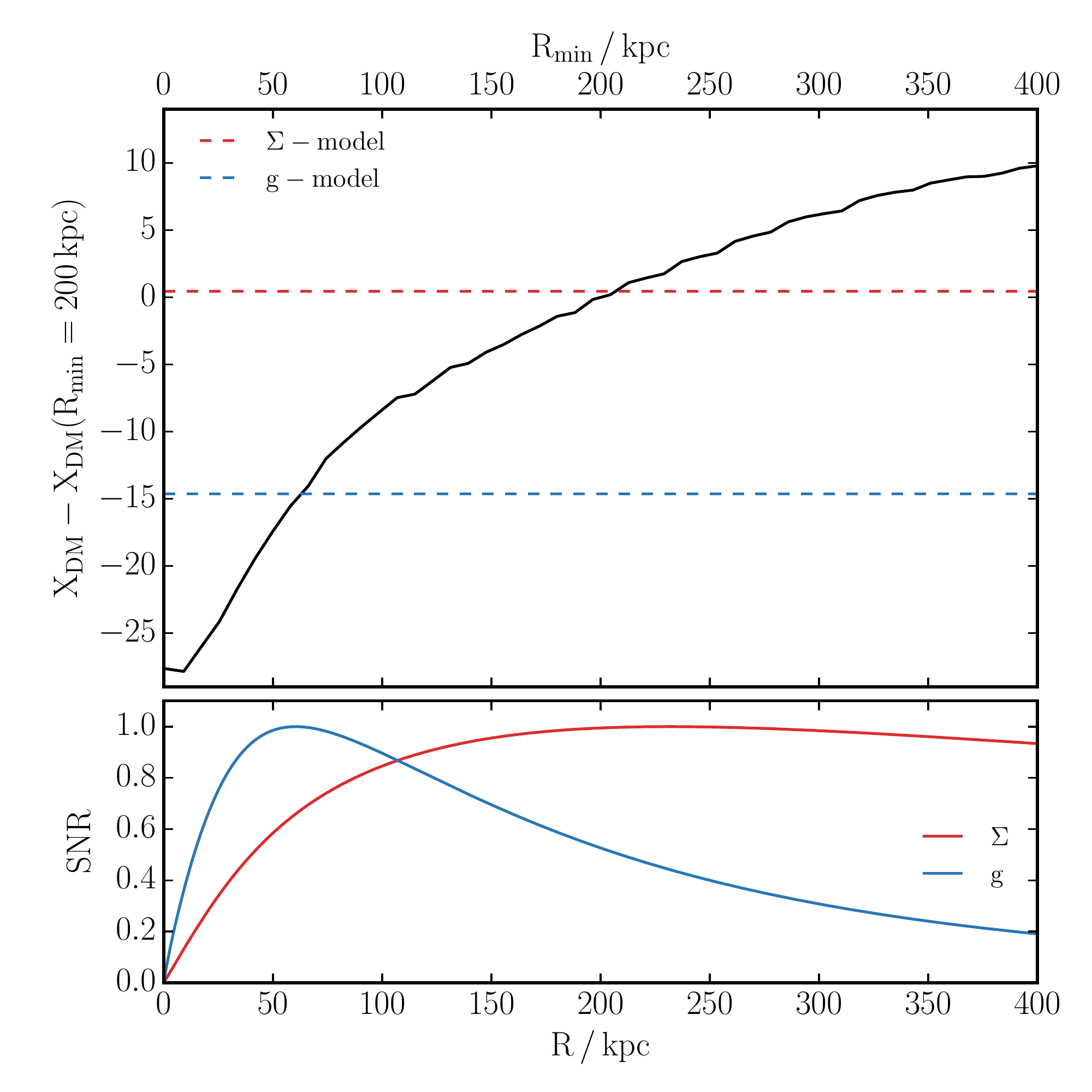}
	\caption{Top panel: the $x$-position of the bullet DM halo measured using shrinking circles on only the DM particles that are part of the bullet halo in the initial conditions. The position is shown as a function of the radius to which the circle is shrunk, with the DM halo shifting to the left as the measurement is made on smaller scales. The best-fit positions of the bullet DM halo from fitting to the projected density and reduced gravitational shear are also shown. Bottom panel: the signal to noise integrated over the projected density or reduced shear map, due to mass within an annulus of fixed width at radius $R$. This was calculated using the projected density as a function of $R$ from the best-fit model to the projected density. Further details are in Appendix~\ref{App:SNR}.}
	\label{fig:halo2_position_2panel}
\end{figure}

In the top row of Fig.~\ref{fig:map_of_halo2} we plot the projected DM distribution, and resulting shear field, at the time of the observed Bullet Cluster, but only using particles that were part of the bullet halo in the initial conditions. What is clear from the projected density is that the mass distribution is not elliptically symmetric, with the peak on small scales being shifted to the left of (i.e. lagging behind) the centre of mass measured on larger scales. This is quantified in the top panel of Fig.~\ref{fig:halo2_position_2panel} where we show the position returned by applying the shrinking circles algorithm on the DM particles from the bullet halo, shrinking down to different final radii, $R_\mathrm{min}$.

The middle and bottom rows of Fig.~\ref{fig:separations_M200match_gas} show the best-fit maps from fitting to the projected surface density and reduced gravitational shear respectively. The projected surface density fit favours a more elliptical halo, centred further to the right, than the shear fit. This, combined with the fact that the halo position shifts left when measuring on smaller scales, suggests that reduced shear is more sensitive to the inner regions of the halo, whereas the projected density fit is more sensitive to larger scales. In the bottom panel of Fig.~\ref{fig:halo2_position_2panel} we show that this is what is expected, plotting (for both $\Sigma$ and $g$) the sum of the signal to noise ratio over the whole map, due to annuli of mass at different radii. The details of this are explained in Appendix~\ref{App:SNR}. We find that this quantity peaks at $R \sim 60 \kpc$ for reduced shear and $R \sim 230 \kpc$ for the projected density, in rough agreement with the shrinking circles $R_\mathrm{min}$ that returns the same position as the respective fitting procedure.

Asymmetry in the DM distribution, and consequent differences in the positions returned by different fitting methods is most pronounced for the collisionless DM case as the cuspy halo is tightly bound to the gas. The formation of DM cores with SIDM reduces the strength of this gravitational binding, such that when the gas is stripped with SIDM it does not drag back the central regions of the DM halo as strongly as with collisionless DM. The stripping of gas is just one mechanism that could cause an asymmetric DM profile, but serves as a cautionary tale for attempts to use offsets between different cluster components to constrain DM's collisional properties. The general result that an asymmetric profile can lead to a measured offset between spatially coincident components, due to them being measured using techniques sensitive to different scales, is an important systematic to consider in future studies.

\subsubsection{Changes to the gas morphology}

Aside from its effect on the DM distribution, the gas itself could potentially be used as a probe of DM self-interactions. Unfortunately, changes to the gas morphology as the DM cross-section is changed are fairly small, with the largest differences being the width and temperature of the shocked region. Increasing the DM cross-section lowers the luminosity-weighted projected temperature in the shocked-gas region, from 30 keV with collisionless DM, to 25 keV with $\sigma / m = 2 \cmsg$, both well within the quoted observational error \citep{2006ESASP.604..723M}.

This decrease in temperature also comes with an increase in the width of the shocked region. The distance between the shock front and the contact discontinuity connecting the shocked gas to the cold gas bullet, increases from $70 \kpc$ with collisionless DM to $110 \kpc$ with $\sigma / m = 2 \cmsg$. While this latter value is in better agreement with the observed distance between the shock front and contact discontinuity ($\sim 140 \kpc$), we find (in agreement with SF07) that this distance is highly sensitive to the concentration of the main halo, making this measurement unsuitable for constraining the DM cross-section.

\section{Conclusions}
\label{sect:conclusions}

We have presented modifications to the \textsc{gadget}-3 code to include elastic DM scattering, allowing us to run simulations with SIDM. We have shown that this code performs as expected when used for simple test cases where the correct behaviour can be predicted analytically. We have also discussed the choice of the numerical parameter, $h$, which is the radius within which particles look for neighbours to scatter with. We have shown that the choice $h \sim \epsilon$ leads to the correct scattering rate within a DM halo, without the computational overhead associated with having an environmentally-dependent $h$.

We have used this code to perform idealised simulations of Bullet Cluster-like systems. With SIDM, the momentum transfer from particles in the main halo to particles in the bullet halo with which they scatter, leads to a tail of scattered particles in the bullet halo that shifts the measured position of this halo relative to the collisionless stars. Our fiducial model for the Bullet Cluster was derived from fits to weak lensing data. Changes to this fiducial model led to changes in the measured offsets between stars and DM, although these changes were small and in a predictable manner.

Our primary conclusion, is that the method used to measure the positions of the different components can have a larger effect than using a different model for the Bullet Cluster. In particular, shrinking circles methods similar to those used by R08, give substantially larger DM--galaxy offsets than more observationally-motivated methods such as parametric fits to the projected density or reduced gravitational shear. This suggests that the  $\sigma / m < 1.25 \cmsg$ constraint placed on the cross-section for DM scattering by R08 is strongly overstated. In fact, for our fiducial model of the Bullet Cluster with $\sigma / m = 2 \cmsg$, the DM--galaxy offset at the time of the observed Bullet Cluster is $\sim 20 \kpc$, which is allowed by the $25 \pm 29 \kpc$ observed offset used by R08 to place their constraint. We produce more robust results by fitting parametric models to the haloes -- which can be done observationally \citep{2005MNRAS.359..417S,2010MNRAS.402L..44R,2012ApJ...757....2G,2015Sci...347.1462H,2015MNRAS.449.3393M,2016ApJ...820...43S}. We recommend that future simulation efforts adopt this, or similarly motivated techniques, to enable a better comparison to observations.

We went on to show results from the first simulations of merging clusters to include both SIDM and gas. The gas does not have much effect on the offset between the stellar and DM components. However, as the gas is stripped it introduces asymmetries into the stellar and DM components, with the central regions of the bullet halo lagging behind the larger-scale centre. This is strongest with collisionless DM where the cuspy halo is tightly bound to the gas. As the methods used observationally to measure the positions of the galaxies and DM will be different, they are likely to be sensitive to different scales. We showed that this can result in a measured offset between these two components even if they have an identical spatial distribution. These asymmetric halo shapes could also be produced by tidal forces or dynamical friction, and these asymmetries are an important potential systematic that could lead to the false detection of SIDM.

\section*{Acknowledgments}
This work was supported by the Science and Technology Facilities Council grant numbers ST/K501979/1 and ST/L00075X/1. RM was supported by the Royal Society. We wish to thank Richard Bower and David Harvey for useful discussions.

This work used the DiRAC Data Centric system at Durham University, operated by the Institute for Computational Cosmology on behalf of the STFC DiRAC HPC Facility (www.dirac.ac.uk). This equipment was funded by BIS National E-infrastructure capital grant ST/K00042X/1, STFC capital 
grants ST/H008519/1 and ST/K00087X/1, STFC DiRAC Operations grant  ST/K003267/1 and Durham University. DiRAC is part of the National E-Infrastructure.

\bibliographystyle{mnras}

\bibliography{cluster_collisions}

\appendix

\section{Tests of Scattering Kinematics}
\label{App:scattering_tests}

\subsection{Scattering Rates}
To test our scattering algorithm, we modelled a uniform cube of particles moving through a constant density background of stationary particles. To allow for simple predictions to be made for the system, we did not allow particles to scatter more than once. All of the particles in the cube had initial velocities $v_{0}$ along the $z$-axis, and gravity was turned off.

The average rate of scattering is $\Gamma = N_\mathrm{c} \, n_\mathrm{b} \, v_{0} \, \sigma_\mathrm{p} $, where $N_\mathrm{c}$ is the number of particles in the cube, and $n_\mathrm{b} = (\rho_\mathrm{b} / m_\mathrm{p})$ is the number density of background particles. This leads to the expected number of interactions after a time $t$
\begin{equation}
N_\mathrm{exp} = N_\mathrm{c} \, \rho_\mathrm{b} \, v_{0} \, (\sigma_\mathrm{p} / m_\mathrm{p}) \, t .
\end{equation}
%
\begin{figure*}
        \centering
        \includegraphics[width=\textwidth]{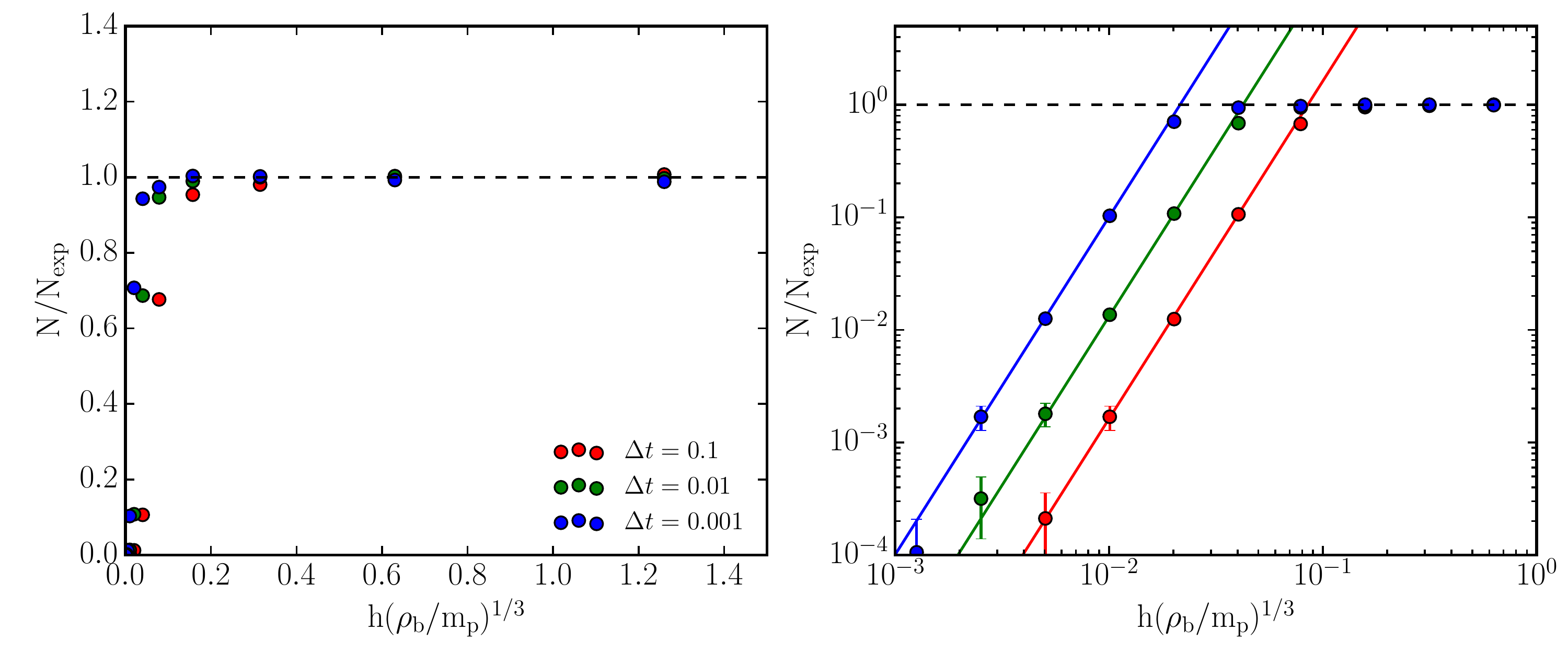}
	\caption{The number of scattering events in our test simulations as a function of neighbour-search radius, $h$. The left panel is similar to Fig.~1 in \citet{Rocha:2013bo}, and we find that we also see a decrease in the rate of scattering, below that expected, when using small $h$. While this happens for $h \lesssim 0.2 (m_\mathrm{p} / \rho_\mathrm{b})^{1/3}$ in agreement with the results in \citet{Rocha:2013bo}, the precise $h$ at which the drop in scattering rate begins is a function of the simulation time step, $\Delta t$. As discussed in the text, this is a result of the probabilities for pairs of particles to scatter within a time step becoming greater than 1. These probabilities are $\propto h^{-3}$, and so in the right panel we show the same data as in the left panel, but plotted on logarithmic scales. The solid lines show $N \propto h^{3}$, the result one expects from probability saturation. For these test simulations, $N_\mathrm{exp} \approx 10^{4}$, and the error bars show the $1 \sigma$ uncertainty, assuming that $N$ is Poisson distributed.}
	\label{fig:test_case_scattered_number}
\end{figure*}

The number of scattering events in these test simulations is plotted in Fig.~\ref{fig:test_case_scattered_number} as a function of the search radius, $h$. For small values of $h$, the number of scattering events falls below that expected. This was noted by \citet{Rocha:2013bo}, who found that scattering was not correctly resolved for $h$ less than 20\% of the mean background inter-particle separation. By running the test case with different time steps, we find that this 20\% is not an intrinsic property of simulating scattering using a Monte Carlo method. Instead we find that the minimum $h$ for which scattering is correctly implemented is a function of the time step, scattering cross-section and the relative velocity of particles.

In general, the scattering rate is insensitive to $h$, as the number of neighbour particles that a particle finds at each time step is proportional to the volume searched ($\propto h^3$), but the probability of scattering from each of those particles follows equation~\eqref{eq:P_ij} ($\propto h^{-3}$). The product of the number of neighbour particles, and the probability of scattering with each of them, gives the total probability of a particle scattering, which does not depend on $h$. This breaks down when the probability to scatter from a neighbour particle becomes greater than unity. At that point, the probability of a particle scattering in a time step is just the probability of finding a neighbour particle during that time step, which goes as $h^3$. For this reason, in this `probability-saturated' regime, the rate of scattering is proportional to $h^3$, as shown by the solid lines in the right panel of Fig.~\ref{fig:test_case_scattered_number}. As the probability for a pair of neighbouring particles to scatter is proportional to $\Delta t / h^{3}$, a smaller $h$ can be used when using shorter time steps.

\subsection{Choosing $h_{si}$}

From equation~\eqref{eq:P_ij} we see that probabilities become greater than unity when
\begin{equation}
h < \left( \tfrac{3}{4 \pi} \sigma_{p} \, |\mathbf{v_{i}} - \mathbf{v_{j}}| \, \Delta t \right)^{1/3}.
\label{eq:prob_sat}
\end{equation}
The time step criterion used for the DM particles in  \textsc{Gadget} is
\begin{equation}
\Delta t = \mathrm{min} \left[ \Delta t_\mathrm{max}, \left( \frac{2 \eta \epsilon}{a}\right)^{1/2} \right],
\label{eq:dt_Gadget}
\end{equation}
where $a$ is a particle's acceleration, $\epsilon$ the gravitational softening, and $\eta$ a dimensionless constant.

As the time steps are dependent on the dynamics of the system being simulated, the constraint on $h$ from equation~\eqref{eq:prob_sat} depends on the content of the simulation. We found that when using the standard time step criterion, the probabilities remained below 1 when $h$ was set equal to the gravitational softening, $\epsilon$. For example, the simulation with our fiducial initial conditions and $\sigma / m = 1 \cmsg$, had a maximum probability for a pair of particles to scatter within a time step of $0.15$, with 99\% of scattering events taking place with a probability $< 0.03$. If running simulations with very large cross-sections, an additional time step criterion could be added to prevent probabilities exceeding unity.

\subsection{Post-scatter kinematics}
\begin{figure}
        \centering
        \includegraphics[width=\columnwidth]{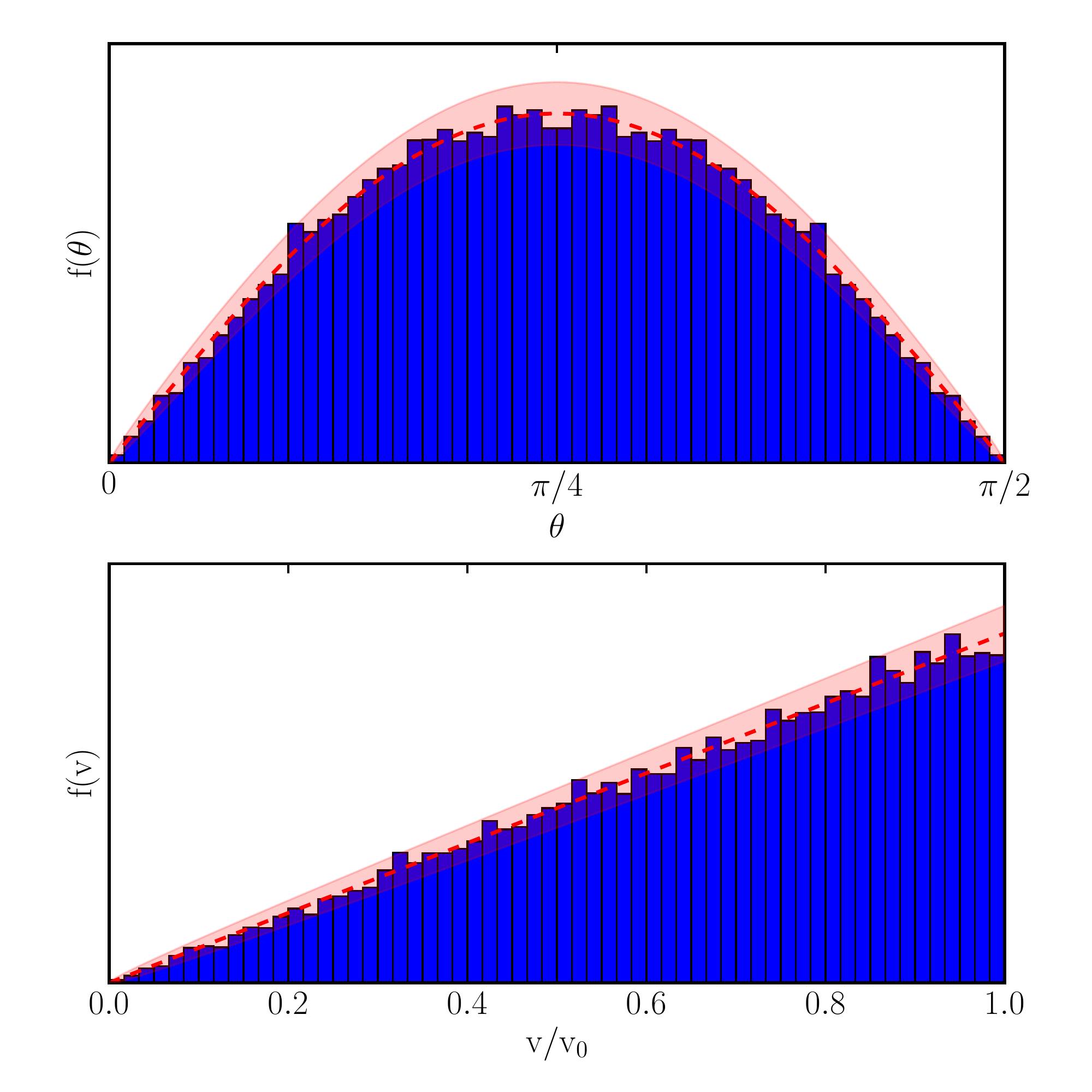}
	\caption{The distribution of polar angles and velocity magnitudes of scattered particles in one of our test simulations. The red dashed line shows the expected distribution, with the red shaded region showing the $2\sigma$ variation about this expectation, assuming the number of particles in each bin is Poisson distributed.}
	\label{fig:test_case_scattered_distributions}
\end{figure}

As well as the rate of scattering, the directions and velocities of scattered particles in our test case were compared to expectations. The expected distribution of velocities and directions is calculated by transforming the differential cross-section from the centre of mass frame of the collisions, into the frame of our simulations. For the case of isotropic scattering, these distributions take on simple forms, with $f(\theta) \propto \sin 2\theta$, and $f(v) \propto v$ for $v \leq v_{0}$, with no particles with velocities greater than $v_{0}$.\footnote{For isotropic scattering, the distribution of scattered particles is the same for those originally part of the background or originally part of the moving cube.} These results are shown in Fig.~\ref{fig:test_case_scattered_distributions}, with results that match expectations.

\section{The contribution of mass at different radii to the projected density and shear signals}
\label{App:SNR}

In Fig.~\ref{fig:halo2_position_2panel} we see that with an asymmetric DM profile, the positions of the halo returned from fitting to the projected density and reduced gravitational shear differ, being similar to the shrinking circles positions of the halo when shrinking to $R_\mathrm{min} = 200$ and $60 \kpc$, respectively. This implies that shear is more sensitive to the central regions of the halo, which appears to be at odds with the maps in Fig.~\ref{fig:map_of_halo2} showing that the projected density (and equivalently convergence) increases towards the centre of the halo, while the shear has a flatter profile. This can be explained by noting that shear is a non-local quantity, and that for a circularly symmetric projected mass distribution the shear at radius $R$ depends on all of the mass within $R$. In fact, the tangential shear ($\gamma_\mathrm{t}$) from a circularly symmetric mass distribution can be written in terms of the `excess surface density'
\begin{equation}
\Delta \Sigma = \bar{\Sigma}(<R) - \Sigma(R) =  \Sigma_\mathrm{crit} \gamma_\mathrm{t}.
\label{excess_surface_density}
\end{equation}

For an annulus of mass at radius $R_0$ with mass $M_0$ the shear internal to $R_0$ is zero, while the shear at $R > R_0$ is simply the average surface density within $R$ divided by the critical surface density. The average surface density is the enclosed mass divided by the area, so
\begin{equation}
\gamma_\mathrm{t} = \frac{1}{\Sigma_\mathrm{crit}} \frac{M_0}{\pi R^2}.
\label{tangential_shear_annulus}
\end{equation}

The noise in the shear map is independent of position, so the signal to noise ratio in a particular pixel is just proportional to the shear there. The number of pixels in an annulus at radius $R$ is proportional to $2 \pi R \, \rmd R$. This implies that the sum of the signal to noise over all pixels in an annulus at $R$ due to the mass $M_0$ at $R_0$ is proportional to $(M_0 / R^2) R \, \rmd R$. Integrating this from $R=R_0$ outwards we find that the sum of the signal to noise over all pixels in the map is proportional to $M_0 \ln (R_\mathrm{max}/R_0)$, where we have truncated the integration at a maximum radius $R_\mathrm{max}$. For $R_\mathrm{max}$ we use half the side length of the square regions used when fitting to shear. As the total signal to noise only grows logarithmically with $R_\mathrm{max}$, this choice is not particularly important.

The mass $M_0$, which is the mass in an annulus at radius $R_0$, is the surface density at radius $R_0$ multiplied by the area of the annulus, so $M_0 \propto \Sigma (R_0) R_0$. As such, the sum of the signal to noise ratio of all pixels in the map due to mass at radius $R_0$ is
\begin{equation}
\mathrm{SNR}_g \propto R_0 \Sigma (R_0) \ln \left( \frac{R_\mathrm{max}}{R_0} \right).
\label{SNR_shear}
\end{equation}

The projected density is a local quantity, leading to the calculation being simpler than for the case of shear. Fitting to the projected density used Poisson statistics, which for large numbers of particles per bin can be approximated by Gaussian statistics. The signal to noise ratio of a single pixel is then $\sqrt{N} \propto \sqrt{\Sigma}$, where $N$ is the number of particles in that pixel. The mass at radius $R_0$ only contributes to the signal at $R_0$, and the number of pixels at radius $R_0$ is proportional to $R_0$. Using this, the sum of the signal to noise ratio of all pixels in the map due to mass at radius $R_0$ is 
\begin{equation}
\mathrm{SNR}_\Sigma \propto R_0 \sqrt{\Sigma (R_0)} .
\label{SNR_projected_density}
\end{equation}

$\mathrm{SNR}_g$ and $\mathrm{SNR}_\Sigma$ are the quantities plotted in the bottom panel of Fig.~\ref{fig:halo2_position_2panel}, where they  have been normalised by their maximum value.

\bsp
\label{lastpage}

\end{document}